\documentclass[a4paper,aps,preprint,nofootinbib]{revtex4}
\pdfoutput=1
\usepackage[latin1]{inputenc}
\usepackage{bbm}
\usepackage{bm}
\usepackage{graphicx}
\usepackage{epsfig}
\usepackage{subfigure}
\usepackage{latexsym}
\usepackage{amsmath}
\usepackage{amsfonts}
\usepackage{amssymb}
\usepackage{wasysym}
\usepackage{color}

\newcommand{\be}{\begin{equation}}
\newcommand{\ee}{\end{equation}}

\newcommand{\bea}{\begin{eqnarray}}
\newcommand{\eea}{\end{eqnarray}}

\definecolor{rossoCP3}{cmyk}{0,.88,.77,.40}
\begin{document}


\title{\textcolor{rossoCP3} {\Large Supersymmetric Extension of Technicolor  \\  \& \\ Fermion Mass Generation}
} 
\author{Matti Antola$ ^{a,b}$\footnote{matti.antola@helsinki.fi}}
\author{Stefano Di Chiara$ ^{c}$\footnote{dichiara@cp3-origins.net}}
 \author{Francesco Sannino $ ^{c}$\footnote{sannino@cp3-origins.net}} 
\author{Kimmo Tuominen $ ^{b,d}$\footnote{kimmo.i.tuominen@jyu.fi}}
\affiliation{\mbox{ $ ^{a}$ Department of Physics, P.O.Box 64, FI-000140, Univ. of Helsinki, Finland} }
\affiliation{\mbox{ $ ^{b}$ Helsinki Institute of Physics, P.O.Box 64, FI-000140, Univ. of Helsinki, Finland}} 
\affiliation{ $ ^{c}$ { \color{rossoCP3}  \rm CP}$^{\color{rossoCP3} \bf 3}${\color{rossoCP3}\rm-Origins} \& the Danish Institute for Advanced Study {\color{rossoCP3} \rm DIAS},\\ 
\mbox{Univ. of Southern Denmark, Campusvej 55, DK-5230 Odense M, Denmark}}
\affiliation{\mbox{ $ ^{d}$ Department of Physics, P.O.Box 35, FI-40014, Univ. of Jyv\"askyl\"a, Finland} }
 

\begin{abstract}
We provide a complete extension of Minimal Walking Technicolor able to account for the Standard Model fermion masses. The model is supersymmetric at energies greater or equal to the technicolor compositeness scale.  We integrate out, at the supersymmetry breaking scale, the elementary Higgses.  We use the resulting four-fermion operators to derive the low energy effective theory. We then determine the associated tree-level vacuum and low energy spectrum properties. Furthermore we investigate the phenomenological viability of the model by comparing its predictions with electroweak precision tests and experimental bounds on the mass spectrum. We then turn to the composite Higgs phenomenology at the LHC and show that current data are already constraining the parameter space of the model. 
\\
{\footnotesize  \it Preprint: CP$^3$-Origins-2011-36 \& DIAS-2011-28}
\end{abstract}
\maketitle

\section{Historical compromise}
\label{Intro}

Dynamical Electroweak (EW) symmetry breaking is arguably one of the most natural extensions of the Standard Model (SM) of particle interactions. The SM Higgs sector is replaced by a gauge theory featuring fermionic matter breaking the SM gauge symmetries dynamically. Models of this type are also known as Technicolor (TC) models since they mimic Quantum Chromodynamics. For recent reviews on dynamical EW symmetry breaking see \cite{Sannino:2009za,Hill:2002ap}. The EW energy scale emerges dynamically within the theory and is stable against quantum corrections. In the SM  the EW scale is directly associated to the quadratic mass operator of the Higgs which is unprotected against quantum corrections leading to unnaturally large fine tuning to keep, order by order in perturbation theory, the EW scale within the experimentally observed value. 

Supersymmetry (SUSY) provides yet another elegant solution to the stabilization of the EW scale. This is achieved by supersymmetrizing the entire SM. By marrying bosons and fermions the symmetries protecting the fermionic sector from acquiring untamed quantum corrections are now felt, via SUSY, also by the scalars of the model and consequently the EW scale   stabilizes. 

Naturalizing the SM comes at a cost both for TC and SUSY. Replacing the Higgs sector with a new strongly coupled theory is, alone, insufficient to give masses to the SM fermions. A new sector is needed which is typically hard to construct  \cite{Ryttov:2011my}, at least if one insists on not having fundamental scalars.  On the other hand, a minimal supersymmetric extension of the SM (MSSM) requires the Higgs sector to be enlarged, introducing a new scale which is, in principle, unrelated to the EW one. This  scale stems out of the unknown scale of the dimensionful coupling ($\mu$) of the operator mixing the two Higgses of the theory.  Furthermore SUSY must break and the correct mechanism of SUSY breaking is still being searched for. In SUSY the advantage is the presence of elementary scalars making it straightforward to give masses to the SM fermions. In this paper we combine the TC and SUSY ideas. 

\vskip .3cm
 
  {\it Why  marrying Technicolor and Supersymmetry?} 

\begin{itemize}

\item[1)]{  Dynamical EW symmetry breaking is already present in the MSSM since QCD breaks the EW symmetry dynamically via a condensate of the quarks. QCD is supersymmetrized at the EW scale.   This is similar to what we plan to do in this work with the difference that the SUSY's Higgses are used for generating the fermion masses and TC drives the bulk of EW symmetry breaking.}

\item[2)]{ This marriage provides a UV complete extension of TC allowing to generate SM fermion masses  in a natural way. }

\item[3)]{ The SUSY extension of TC allows to decouple the scale of SUSY from the EW one, thus reducing the tension with the LHC constraints for SUSY.} 


\end{itemize}

 In this work we will investigate a specific SUSY extension of Minimal Walking Technicolor (MWT) \cite{Antola:2010nt,Sannino:2004qp,Dietrich:2005jn,Foadi:2007ue} \footnote{Minimal walking models emerged while investigating the phase diagram of strongly coupled theories as a function of the number of flavors, colors and matter representation \cite{Sannino:2004qp,Dietrich:2006cm,Ryttov:2007sr,Ryttov:2007cx,Frandsen:2010ej,Mojaza:2010cm,Pica:2010mt,Antipin:2009wr,Pica:2010xq,Ryttov:2010iz}. See also \cite{Sannino:2009za} for a more complete list of references.}. We dubbed this model Minimal Supersymmetric Conformal Technicolor (MSCT). 
Here we show that MSCT passes the EW constraints, it does well with respect to Flavor Changing Neutral Currents (FCNC), and finally the spectrum of the model is rich, phenomenologically viable and testable at the LHC.

 In Section \ref{MWTs} we briefly summarize the salient features of MWT and set the stage for going beyond technicolor.
  We introduce in Section \ref{UVCTs} a supersymmetrized version of MWT able to generate the fermion masses. We then determine the low energy effective theory at the electroweak scale. In Section \ref{SPQR} we determine the vacuum properties, the EW precision parameters, and the spectrum. We then make a scan of the parameter space of the theory. The composite Higgs phenomenology of the theory is studied in Section \ref{LHCPh}. A different regime of the theory is studied in Section \ref{LSSs}. We conclude in Section \ref{conclusione}.  Finally we added Appendix \ref{appn4} to summarize the ${\cal N}=4$ Lagrangian and Appendix \ref{AppA} to show the scalar mass matrices.

\section{Minimal Walking Technicolor Review}
\label{MWTs}
We start from a TC model featuring a minimal particle spectrum required to break correctly EW. This is necessary to keep the $S$ parameter small, as required by the experimental bounds, since the naive value of $S$ associated with the techniquark spectrum is proportional to the number of EW doublets.

Moreover, when considering generic extensions of TC  the resulting theory should  yield the correct top quark mass without generating large flavor changing neutral currents.  One possibility is to devise models in which the TC coupling  walks \cite{Holdom:1981rm,Appelquist:1986an,Yamawaki:1985zg}, as opposed to QCD-like running. A model encompassing minimality as well as walking behavior is MWT \cite{Sannino:2004qp,Dietrich:2005jn}. The MWT extension of the SM has the following gauge group:
\be
SU(2)_{TC}\times SU(3)_C \times SU(2)_L \times U(1)_Y
\ee
while its particle spectrum features all the SM particles besides the Higgs scalar, plus an EW techniquark doublet in the adjoint representation of $SU(2)_{TC}$ as well as its right-handed components:
\be Q_L^a=\left(\begin{array}{c} U^{a} \\D^{a} \end{array}\right)_L , \qquad U_R^a \
, \quad D_R^a \ ,  \qquad a=1,2,3 \ ,\ee with $a$ being the adjoint color index of SU(2). The left-handed fields are arranged in three
doublets of the SU(2)$_L$ weak interactions in the standard fashion. The condensate is $\langle \bar{U}U + \bar{D}D \rangle$ which
correctly breaks the EW symmetry, and therefore gives mass to the $W^\pm$ and $Z$ bosons.
A new weakly charged fermionic doublet which is a TC singlet \cite{Dietrich:2005jn} is also added to cancel the Witten topological anomaly: 
\be L_L =
\left(
\begin{array}{c} N \\ E \end{array} \right)_L , \qquad N_R \ ,~E_R \ . \ee 
The following hypercharge assignment is anomaly-free and yields integer electromagnetic (EM) charges:
\begin{align}
Y(Q_L)=&\frac{1}{2} \ ,&\qquad Y(U_R,D_R)&=\left(1,0\right) \ , \\
Y(L_L)=& -\frac{3}{2} \ ,&\qquad
Y(N_R,E_R)&=\left(-1,-2\right) \ \label{assign2} .
\end{align}

To discuss the symmetry properties of the
theory it is
convenient to use the Weyl basis for the fermions and arrange them in the following vector that transforms according to the
fundamental representation of SU(4):
\be \eta= \begin{pmatrix}
U_L \\
D_L \\
-i\sigma^2 U_R^* \\
-i\sigma^2 D_R^*
\end{pmatrix},
\label{SU(4)multiplet} \ee where $U_L$ and $D_L$ are the left-handed techniup and technidown, respectively, $U_R$ and $D_R$ are
the corresponding right-handed particles, and $\sigma^2$ is a Pauli matrix. Assuming the standard
breaking to the maximal diagonal subgroup, the SU(4) symmetry
spontaneously breaks to $SO(4)$. Such a breaking is driven by the
following condensate \be \langle \eta_i^\alpha \eta_j^\beta
\epsilon_{\alpha \beta} E^{ij} \rangle =-2\langle \overline{U}_R U_L
+ \overline{D}_R D_L\rangle \ , \label{conde}
 \ee
where the indices $i,j=1,\ldots,4$ denote the components
of the tetraplet of $\eta$, and the Greek indices indicate the ordinary
spin. The matrix $E$ is a $4\times 4$ matrix defined in terms
of the 2-dimensional unit matrix as
 \be E=\left(
\begin{array}{cc}
0 & \mathbbm{1} \\
\mathbbm{1} & 0
\end{array}
\right) \ . \ee

Here $\epsilon_{\alpha \beta}=-i\sigma_{\alpha\beta}^2$ and $\langle
 U_L^{\alpha} {{U_R}^{\ast}}^{\beta} \epsilon_{\alpha\beta} \rangle=
 -\langle  \overline{U}_R U_L
 \rangle$. A similar expression holds for the $D$ techniquark.
The above condensate is invariant under an $SO(4)$ symmetry. This leaves us with nine broken  generators with associated Goldstone bosons.

The Lagrangian of MWT,  for a general anomaly-free hypercharge assignment, has been constructed in \cite{Foadi:2007ue}. There is no vacuum alignment problem for this model \cite{Dietrich:2009ix}. 

\begin{figure}[h]
\begin{center}
\includegraphics[width=18pc]{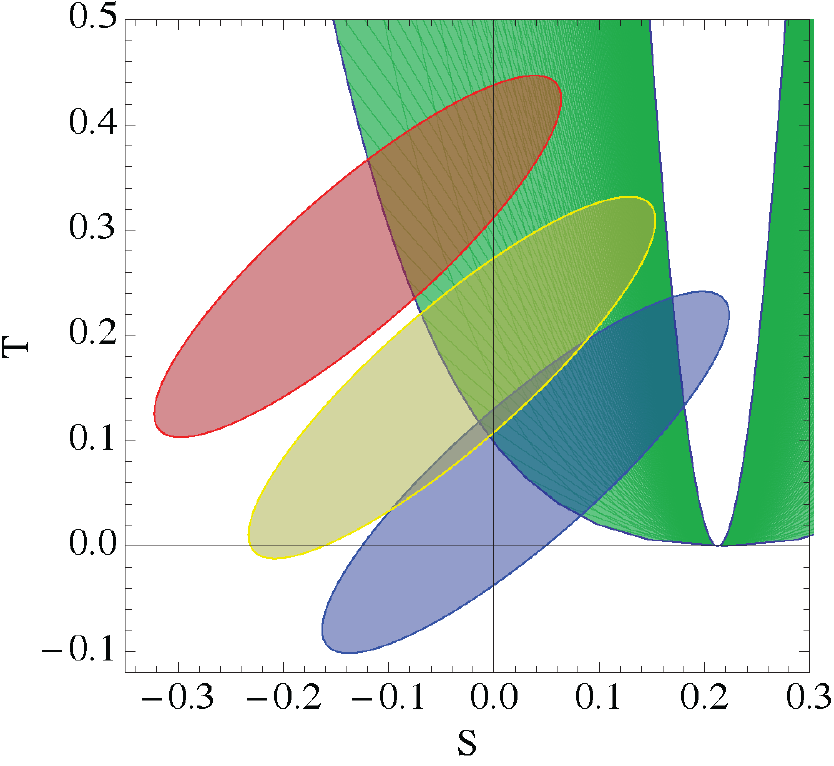}
\caption{\label{viaST}The ellipses represent the 90\% confidence region for the S and T parameters, and are obtained, from lower to higher, for a reference Higgs mass of 117 GeV, 300 GeV, and 1 TeV, respectively. The contribution from the MWT theory as a function of the $N$ and $E$ lepton masses is expressed by the green region, with $m_Z \leqslant m_{N,E}\leqslant 10 m_Z$.}
\vspace{1pc}
\end{center}
\end{figure}

The contribution of the leptonic heavy doublet allows MWT to satisfy the experimental constraints on the $S$ and $T$ parameters. This is seen from Fig.(\ref{viaST}), where we have plotted the theoretical prediction of MWT for the EW precision parameters, given by $S_{\mathrm{naive}}\simeq 1/2\pi$ plus a perturbative contribution of the heavy leptons to $S$ as well as to $T$ \cite{He:2001tp}. A general discussion of the constraints on the precision tests beyond $S$ and $T$ has been investigated in \cite{Foadi:2007se}, while the unitarity problem for strongly interacting models of dynamical EW symmetry breaking has been investigated in \cite{Foadi:2008ci,Foadi:2008xj}. The investigation of the effects of conformal dynamics on the $S$ parameter appeared in \cite{Sannino:2010ca,Sannino:2010fh,DiChiara:2010xb,Sondergaard:2011ps}.  The new bound on $S$ \cite{Sannino:2010ca,Sannino:2010fh,DiChiara:2010xb} is consistent with the estimate used above.

\subsection{Effective Lagrangian for MWT before addressing the fermion mass generation}
\label{effMWT} 

The scalar and pseudoscalar degrees of freedom replacing the SM Higgs sector at the EW scale consist, in our model, of a composite Higgs and its pseudoscalar partner, as well as nine pseudoscalar Goldstone bosons and their scalar partners. These
can be assembled in the matrix
\begin{eqnarray}
M = \left[\frac{\sigma+i{\Theta}}{2} + \sqrt{2}(i\Pi^a+\widetilde{\Pi}^a)\,X^a\right]E \ ,
\label{M}
\end{eqnarray}
which transforms under the full SU(4) group according to
\begin{eqnarray}
M\rightarrow uMu^T \ , \qquad {\rm with} \qquad u\in {\rm SU(4)} \ .
\end{eqnarray}
The $X^a$'s, $a=1,\ldots,9$ are the generators of the SU(4) group which do not leave  the vacuum expectation value (VEV) of $M$ invariant
\begin{eqnarray}
\langle M \rangle = \frac{v_w}{2}E
 \ .
\end{eqnarray}
Note that the notation used is such that $\sigma$ is a \emph{scalar}
while the $\Pi^a$'s are \emph{pseudoscalars}. It is convenient to
separate the fifteen generators of SU(4) into the six that leave the
vacuum invariant, $S^a$, and the remaining nine that do not, $X^a$.
Then the $S^a$ generators of the SO(4) subgroup satisfy the relation
\begin{eqnarray}
S^a\,E + E\,{S^a}^{T} = 0 \ ,\qquad {\rm with}\qquad  a=1,\ldots  ,  6 \ ,
\end{eqnarray}
so that $uEu^T=E$, for $u\in$ SO(4). The explicit realization of the generators is shown in the appendix of \cite{Foadi:2007ue}. The matrix $M$ is invariant in form under U(4)$\equiv$SU(4)$\times$U(1)$_{\rm
A}$, rather than just SU(4). However the U(1)$_{\rm A}$ axial symmetry is anomalous, and is therefore broken at the quantum level. \footnote{We used a linear sigma model to describe some of the low lying states. A well known theorem due to Haag \cite{9166} demonstrated  the equivalence, at tree level, of different effective Lagrangians with the same number of asymptotic fields respecting the same global symmetries. We have also checked this directly by comparing the linear Lagrangian with the nonlinear one  in \cite{Foadi:2008xj}.}

The connection between the composite scalars and the underlying techniquarks can be derived from the transformation properties under SU(4), by observing that the elements of the matrix $M$ transform like techniquark bilinears:
\begin{eqnarray}
M_{ij} \sim \eta_i^\alpha \eta_j^\beta \varepsilon_{\alpha\beta}  =\eta_i^\alpha {\eta_j}_{\alpha} \quad\quad\quad {\rm with}\ i,j=1\dots 4.
\label{M-composite}
\end{eqnarray}
 The effective linearly transforming Lagrangian up to $SU(4)$ breaking terms, and prior to introducing a flavor complete extension reads:
 \be
{\cal L}_{MWT} = \frac{1}{2}\text{Tr}\left[DM^{\dagger}DM\right]-\mathcal{V}_{M}
\,,
\label{LMWT}
\ee
where the covariant derivative is given by
\[
DM=\partial M-i\left[GM+MG^T\right]\;,
\]
with
\[
G=g_W W^a L^a+g_Y B Y_M\\
\]
and
\[
L^a=\left(\begin{array}{cc}
\frac{\sigma^{a}}{2} & 0\\
0 & 0\end{array}\right)\,,\qquad
Y_M={\textrm{diag}}\left(\frac{1}{2},\frac{1}{2}, -1,0\right).
\]
The SU$(4)$  preserving effective potential is
\be
\mathcal{V}_{M}=
-\frac{m^{2}}{2}\text{Tr}\left[M^{\dagger}M\right]
+\frac{\lambda}{4}\text{Tr}\left[M^{\dagger}M\right]^{2}
+\lambda^{'}\text{Tr}\left[M^{\dagger}MM^{\dagger}M\right]
-2\lambda^{''}\left[\det M+\det M^{\dagger}\right]  \ .
\label{su4} 
\ee
One can also add vector resonances to this Lagrangian as done in \cite{Foadi:2007ue}. $SU(4)$ breaks spontaneously to $SO(4)$  for positive $m^2$.  Stability of the potential furthermore requires
\be
\lambda>0,\ \lambda '>0,\ \lambda+\lambda '>\lambda ''>0.
\label{stab}
\ee

Three of the nine Goldstone bosons associated with the broken generators become the longitudinal degrees of freedom of
the massive weak gauge bosons, while the extra six Goldstone bosons will acquire a mass due to Extended TC (ETC) interactions as well as the EW interactions per se.

The potential ${\cal V}(M)$ is SU(4) invariant. It produces a VEV
which parameterizes the techniquark condensate, and spontaneously
breaks SU(4) to SO(4). In terms of the model parameters the VEV is
\begin{eqnarray}
v_w^2=\langle \sigma \rangle^2 = \frac{m^2}{\lambda + \lambda^\prime - \lambda^{\prime\prime} } \ ,
\label{VEV}
\end{eqnarray}
while the composite Higgs mass 
\begin{eqnarray}
M_H^2 = 2\ m^2 \ .
\end{eqnarray}
The linear combination $\lambda + \lambda^{\prime} -
\lambda^{\prime\prime}$ corresponds to the composite Higgs self coupling in
the SM. The three pseudoscalar mesons $\Pi^\pm$, $\Pi^0$ correspond
to the three massless Goldstone bosons which are absorbed by the
longitudinal degrees of freedom of the $W^\pm$ and $Z$ boson. The
remaining six uneaten Goldstone bosons are technibaryons, which are massless at this stage, 
\begin{eqnarray}
M_{\Pi_{UU}}^2 = M_{\Pi_{UD}}^2 = M_{\Pi_{DD}}^2 =0 \ .
\label{mwtgoldstonemass}
\end{eqnarray}
However, all these are expected to acquire tree-level  masses through not yet specified ETC interactions \cite{Foadi:2007ue}. 
We will explicitly show in Sec. \ref{SPQR} that after supersymmetrizing MWT, no unacceptably light exotic particles appear in the low energy spectrum.

The remaining scalar and pseudoscalar masses are
\begin{eqnarray}
M_{\Theta}^2 & = & 4 v_w^2 \lambda^{\prime\prime} \nonumber \\
M_{A^\pm}^2 = M_{A^0}^2 & = & 2 v_w^2 \left(\lambda^{\prime}+\lambda^{\prime\prime}\right)
\end{eqnarray}
for the technimesons, and
\begin{eqnarray}
M_{\widetilde{\Pi}_{UU}}^2 = M_{\widetilde{\Pi}_{UD}}^2 = M_{\widetilde{\Pi}_{DD}}^2 =
  2 v_w^2 \left(\lambda^{\prime} + \lambda^{\prime\prime }\right) \ ,
\end{eqnarray}
for the technibaryons. 
To gain insight into some of the mass relations one can use the results and ideas introduced in \cite{Hong:2004td,Dietrich:2009ix}. Furthermore recent lattice investigations for MWT and next to MWT \cite{Catterall:2007yx,Hietanen:2008mr,DelDebbio:2008zf,Catterall:2008qk,Shamir:2008pb,DeGrand:2008kx,Fodor:2009wk,Fodor:2009ar,Hietanen:2009az,Bursa:2009we,DelDebbio:2009fd,DelDebbio:2010hx,DelDebbio:2010hu,DeGrand:2010na,Shamir:2010cq,Kerrane:2010xq,Kogut:2010cz,Sinclair:2010kf,Sinclair:2010be,Fodor:2011tw,DeGrand:2011qd,Kogut:2011ty} are providing relevant information on the dynamics of  MWT, before coupling it to the SM and without extending the theory to accommodate for the SM fermion masses.  
  
MWT predictions for LHC have been first investigated in \cite{Belyaev:2008yj} while the most recent LHC constraints on the parameter space have been provided in \cite{Andersen:2011nk}. A more general and recent analysis of TC model phenomenology and model classification relevant for LHC searches can be found in \cite{Andersen:2011yj}.  Other interesting analyses related to TC phenomenology have also appeared \cite{Banerjee:2011nq,Chivukula:2011ue}.  

\subsection{Need to go beyond Technicolor}
\label{BT}

The analysis above shows that if the dynamics is such that  $SU(4)$ breaks to $SO(4)$ then one can correctly break the EW symmetry dynamically, provided the embedding of the SM gauge symmetries is the one envisioned above. However some of the composite states remain massless. Furthermore the SM fermions have not yet acquired mass. This situation is common to generic TC extensions of the SM and demonstrates the need of yet new dynamics at some higher energy scale to complete the model. In this framework not even the new leptons have acquired a mass. Therefore we cannot yet compute corrections to precision observables in a consistent manner. More generally the needed extension can also trigger or modify the pattern of chiral symmetry breaking as we shall show below.

The minimal approach used in \cite{Foadi:2007ue} has been to recouple the composite Higgs sector of MWT to the SM fermions in a phenomenological way. It was assumed there that an ETC dynamics existed, leading, at the EW scale, to an extension of the effective Lagrangian able to give masses to the Goldstone bosons and as well as to the SM fermions.  While this approach is phenomenologically justified, it is important to demonstrate the existence of {\it at least} one natural ultraviolet completion of MWT solving the generation of the SM fermion masses at a more fundamental level. 

Different extensions of TC have been proposed to address the SM fermion masses problem:

\begin{itemize}

\item The most radical approach is to assume that there are no fundamental scalars in Nature. This ambitious program requires the existence of one or more new strongly interacting theories featuring only new matter fermions and gauge bosons beyond the TC one \cite{Eichten:1979ah}. Progress in this direction has been made by Ryttov and Shrock in a series of recent papers \cite{Ryttov:2011my,Chen:2010er,Ryttov:2010fu,Ryttov:2010hs,Ryttov:2010jt,Ryttov:2010kc}. 

\item Another approach is that of bosonic TC \cite{Simmons:1988fu, Kagan:1991gh,Carone:1992rh, Carone:1994mx, Antola:2009wq}. In these theories at least a massive scalar with Yukawa couplings to both techniquarks and SM fermions is introduced. 

\item Another possibility, the one we consider here, is to introduce SUSY above the TC scale.  In this scenario SUSY works as an ETC model linking the TC sector to the SM fermions. It is worth noting that, in addition to the top mass, also neutrino masses are problematic in TC extensions. The present approach addresses the first problem while alleviates the second. Furthermore, the FCNCs related to slepton and squark mass matrices are suppressed since the SUSY scalars are taken to be typically heavy with respect to the EW scale. Since TC drives EW symmetry breaking it solves the little hierarchy problem of SUSY by allowing the elementary Higgs tree-level masses to be at the SUSY breaking scale \cite{Dine:1981za, Dine:1990jd, Dobrescu:1995gz,Murayama:2003ag,Antola:2010nt,Evans:2010ed,Gherghetta:2011na}.

\end{itemize}
 
\section{Supersymmetrizing MWT Generates Fermion Masses}
\label{UVCTs}

We explore in this work a minimal supersymmetric extension of MWT with the hypercharge assignment given in Eq.(\ref{assign2}). The model features the scalar superpartners for $Q_L,\ U_R,\ L_L,\ N_R,\ E_R$ but not for $D_R$, which plays the role of the technigaugino. These fields are assembled in the superfields presented with the corresponding quantum numbers in Table~\ref{MSCTsuperfields}. The rest of the particle spectrum is that of MSSM, with $H_u$ and $H_d$   the $Y=\pm1/2$ Higgs superfields.  
\begin{table}[h]
\centering
\begin{tabular}{c|c|c|c|c}
Superfield & SU$(2)_{TC}$ & SU$(3)_{\text{c}}$ & SU$(2)_{\text{L}}$ & U$(1)_{\text{Y}}$\tabularnewline[\doublerulesep]
\hline \hline
$\Phi_{1,2}$ & Adj & $1$ & $\square$ & 1/2\tabularnewline[\doublerulesep]
$\Phi_{3}$ & Adj & 1 & 1 & -1\tabularnewline[\doublerulesep]
$V$ & Adj & 1 & 1 & 0\tabularnewline[\doublerulesep]
$\Lambda_{1,2}$ & 1 & 1 & $\square$ & -3/2\tabularnewline[\doublerulesep]
$N$ & 1 & 1 & 1 & 1\tabularnewline[\doublerulesep]
$E$ & 1 & 1 & 1 & 2\tabularnewline[\doublerulesep]
\end{tabular}\caption{MSCT ${\cal N}=1$ superfields}
\label{MSCTsuperfields}
\end{table}

At the {\it microscopical} level the full theory, which is renormalizable, is defined by the canonical kinetic terms and the following superpotential\footnote{In the superpotential we impose R-parity invariance, since it allows the model to satisfy more easily the experimental bounds from accelerators. However, there are additional, R-parity violating terms that can in principle be added to the superpotential  in Eq.\eqref{spmwt}.}:
\be
P=P_{MSSM}+P_{TC},
\label{spot}
\ee
where $P_{MSSM}$ is the MSSM superpotential\footnote{For a thorough review of MSSM see \cite{Martin:1997ns} and references therein.}, and
\be
P_{TC}=-\frac{g_{TC}}{3\sqrt{2}} \epsilon_{ijk} \epsilon^{abc} \Phi^a_i \Phi^b_j \Phi^c_k+y_U \epsilon_{ij3}\Phi^a_i {{H_u}_j}\Phi^a_3+y_N \epsilon_{ij3}\Lambda_i {H_{u}}_j N+y_E \epsilon_{ij3}\Lambda_i {H_{d}}_j E+y_R \Phi^a_3 \Phi^a_3 E.
\label{spmwt}
\ee
Here $g_{TC}$ is the TC gauge coupling. The first observation is that the supersymmetrized version of the TC theory in isolation corresponds to ${\cal N} =4$ super Yang-Mills and therefore its beta function vanishes to all orders. It is for this reason, and for the minimality of the particle spectrum, that we have termed this model Minimal Supersymmetric Conformal TC (MSCT) \cite{Antola:2010nt} \footnote{A more recent example of supersymmetric conformal technicolor model is presented in \cite{Evans:2010ed, Azatov:2011ht}. This model, however, is not minimal since {\it junk fields} are introduced to make the technicolor sector conformal.}. The SUSY breaking sector as well as the rest of the Lagrangian follows straightforwardly \cite{Antola:2010nt} from Table~\ref{MSCTsuperfields} and Eqs.(\ref{spot}), (\ref{spmwt}). As shown in \cite{Antola:2010nt} the embedding of the MSSM within the ${\cal N}=4$ theory breaks it to ${\cal N}=1$.

This model constitutes our fundamental description in terms of the elementary degrees of freedom and forces. The  relevant scales of the problem are the supersymmetric breaking scale $m_\textrm{SUSY}$ and the EW one which we identify with the low energy strongly coupled regime of the TC theory, $\Lambda_{TC} \sim 4 \pi v_w$, which for $v_{w}\simeq 246$~GeV implies $\Lambda_{TC} \sim 3$~TeV. We will consider here the following order
\begin{equation}
m_\textrm{SUSY} > \Lambda_{TC} \ .
\end{equation}
In practical numerical estimates we enforce the last relation by simply requiring $m_\textrm{SUSY}>5$ TeV. With this order the EW symmetry is broken dynamically. However, having arranged the order of these scales does not automatically imply the knowledge of  the supersymmetric particle spectrum pattern. We arrange the spectrum in the following way: 

\begin{itemize} 

\item[1)] The soft scalar masses of the fundamental scalars are taken to be of the order of $m_\textrm{SUSY}$.

\item[2)] The gaugino masses are also taken to be of the order of $m_\textrm{SUSY}$ with the exception of the technigaugino mass $M_D$ taken to be lighter than $m_\textrm{SUSY}$. As we will show in section \ref{SPQR}, a relatively small Majorana mass for $D_R$, nearly degenerate with the mass of $U$, is necessary to have a small $T$ parameter, as required by the EW precision data.  Moreover, if $D_R$ was taken to be very heavy compared to $\Lambda_{TC}$, the
low energy theory would be different from the one (MWT) we are using in the
present work.

\item[3)]  We will argue that in our model the $\mu$ parameter, which gives the mass of the Higgsinos, is much larger than 
$\Lambda_{\textrm{TC}}$. Therefore the Higgsinos are ignored in low energy phenomenology.

\item[4)] The remaining states acquire a mass, at most, of the order of $\Lambda_{TC}$. 

\end{itemize}
Since the fundamental scalars are taken to be heavy, i.e. with masses of the order of $m_\textrm{SUSY}$, there is much less fine tuning than in MSSM. Also note that the $\mu$ problem in this model is less severe than in
the MSSM, given that the SUSY breaking scale is higher than in the MSSM.
In addition, in the end of this section we will give an example of a
parameter space region where $\mu$ is much larger than the SUSY breaking scale.

Next we quantify how much the TC extension envisioned here, able to give mass to the SM fermions, modifies the vacuum, spectrum, and dynamics of the original TC model while still allowing for proper dynamical EW symmetry breaking.  The modification of the TC dynamics and consequently of its phenomenology is substantial and cannot be neglected. A similar large effect on the TC dynamics was investigated in \cite{Fukano:2010yv} and led to the discovery of  {\it ideal walking} models.

\subsection{Integrating out SUSY Fields}
\label{eLs}

We use the general soft SUSY breaking Lagrangian written explicitly in \cite{Antola:2010nt}. We assume that all the soft mass parameters are degenerate with common scale $m_\textrm{SUSY}$ except for the technigaugino $\bar{D}_R$ mass term which is taken to be light, as explained above. This will induce a modification of the low energy effective theory describing the composite Higgs dynamics, schematically:
 \be
{\cal L }_{MWT} \rightarrow {\cal L}_{MWT}+{\cal L}_{ETC} \ .
\ee
We determine ${\cal L}_{ETC}$ by first integrating out the scalar Higgses and technisquarks. To determine this effective Lagrangian we need the following microscopic MSCT Lagrangian terms:
\bea
-\mathcal{L}_{\text{Y,MSCT}} & = & \tilde{H}_u \cdot F_u + \tilde{H}_d \cdot  F_d+\textrm{h.c.} \\
F_u & = & q_{Lu}^i Y_u ^{i}\bar{u}_R^i+y_U Q_L \bar{U}_R+y_N L_L \bar{N}_R \\ 
F_d & = & q_{Ld}^i Y_d^{i}\bar{d}_R^i+l_L^i Y_l^{i}\bar{e}_R^i+y_E L_L \bar{E}_R \ .
\label{Lmic}
\eea
The fields $\tilde{H}_u$ and $\tilde{H}_d$ are the MSSM Higgses\footnote{We denote the scalar components of a chiral supermultiplet with a tilde.}. The flavor index is denoted by $i=1...3$ and it is  summed over. The matrices $Y_u$, $Y_d$, and $Y_l$ are diagonal, and the CKM matrix $V$ is hidden in the definitions of the vectors 
\begin{equation}
q_{Lu}^i=(u_L^i,V^{ij}d_L^j) \ , \qquad {\rm and} \qquad q_{Ld}^i=(V^{\dagger ij} u_L^j,d_L^i) \ .
\end{equation}
 Also, the contraction between SU$(2)_L$ doublets with the antisymmetric two-index tensor $\epsilon$ is indicated by a dot symbol ($\cdot$). Furthermore, the Yukawa interaction between technisquarks and techniquarks stemming from superpotential and gauge interactions is given in Eq.\eqref{n4yukawa}.

The MSSM Higgs potential is
\be
\label{vin}
V_\textrm{MSSM}=\left(m_\textrm{SUSY}^2+|\mu|^2\right) |\tilde{H}_u|^2+\left(m_\textrm{SUSY}^2+|\mu|^2\right) |\tilde{H}_d|^2-\left(b\tilde{H}_u\tilde{H}_d+\textrm{h.c.}\right)+...
\ee
The Higgs states are diagonalized via:
\bea
\left(\begin{array}{c}
\tilde{H}_u\\
\tilde{H}^c_d\\
\end{array}\right)
& = &
\frac{1}{\sqrt{2}}\left(\begin{array}{cc}
1 & - 1  \\
1  & 1  \\
\end{array}\right)
\left(\begin{array}{c}
\tilde{H}_1\\
\tilde{H}_2\\
\end{array}\right) \ ,
\eea
where $\tilde{H}^c_d=\epsilon \tilde{H}^*_d$. Their tree-level physical masses $m_1^2=\mu ^2+m_{\text{SUSY}}^2-b$ and $m_2^2=\mu ^2+m_{\text{SUSY}}^2+b$ are traded for two convenient parameters $\theta$ and $m_s$ as follows:
\be
\frac{1}{2} \left(\frac{1}{m_1^2}+\frac{1}{m_2^2}\right)=\frac{c_{\theta
   }^2}{m_{s}^2}\quad,\quad \frac{1}{2}
   \left(\frac{1}{m_1^2}-\frac{1}{m_2^2}\right)=\frac{c_{\theta } s_{\theta
   }}{m_{s}^2}\ .
\label{sincos}
\ee
In terms of the original potential parameters, Eq.\eqref{vin}, we have
\be
m^2_s=\left(\mu ^2+m_{\text{SUSY}}^2\right)\frac{(\mu ^2+m_{\text{SUSY}}^2)^2-b^2}
{\left(\mu ^2+m_{\text{SUSY}}^2\right)^2+b^2}\quad,\quad
t_\theta=\frac{b}{\mu ^2+m_{\text{SUSY}}^2}\;,
\label{msparameters}
\ee
We use the notation $c_\theta,s_\theta,t_\theta$ for, respectively, $\cos\theta,\sin\theta,\tan\theta$. 
Decoupling the heavy scalars leads to the following fermion Lagrangian:
\bea
{\cal L}_{4fermi}&=&\frac{c_\theta^2}{m_s^2}\left(F_u^\dagger F_u+F_d^\dagger F_d\right)-\frac{c_\theta s_\theta}{m_s^2}\left(F_u\cdot F_d+\textrm{h.c.}\right)\nonumber\\
&&- \frac{1}{2} M_D \left(D_R D_R+\textrm{h.c.}\right)+\frac{g_{\text{TC}}^2}{m_\textrm{SUSY}^2} \epsilon_{abc}\epsilon_{cde}\eta_i^{\alpha a}{\eta_{j\alpha}^{ b}} \eta^{*d}_{i\beta} \eta^{*\beta e}_j \ ,
\label{4fL}
\eea
with the techniquarks $\eta$ defined in Eq.\eqref{SU(4)multiplet}. We have retained the operators in mass dimension less or equal to six. We have that $i$ and $j$ indicate the $SU(4)$ flavors, the first letters of the alphabet are reserved for the adjoint $SU(2)$ technicolor indices, while the Greek indices are for $SL(2,C)$ spinors. The color indices are contracted and suppressed, while the TC indices, running from 1 to 3, are written explicitly only in the last term. Note that these terms do not contribute to e.g. $K-\bar{K}$ mixing at tree level, as a consequence of the unitarity of the CKM matrix.

 \subsection{Effective theory below $\Lambda_{TC}$}
The SM fermion masses as well as the fourth lepton family ones  arise from the following four-fermion operator
\be
\eta^{T}Z\eta \ , 
\ee 
with
\be 
Z_{ij}=\frac{y_Uc_\theta\omega}{m_s^2}\left[\delta_{ik}c_\theta\left( q_{Lu}^{*k} Y_u^{*}\bar{u}_R^{*}
+ y_N^* L_{L}^{*k} \bar{N}_R^*\right)
- \epsilon_{ik} s_\theta\left( q_{Ld}^k Y_d \bar{d}_R
+  l_{L }^kY_l \bar{e}_R
+ y_E L_{L}^k \bar{E}_R \right) \right]\delta_{3,j} \ ,
\label{Zsp}
\ee
upon condensation of the techniquarks\footnote{Notice that the indices $i$ and $j$ in Eq.\eqref{Zsp} run from 1 to 4, while $k=1,2$ is the weak isospin.}. The spurion $Z$ transforms as $Z\rightarrow u^* Z u^\dagger$ under SU$(4)_R$.  Having derived the four-fermion theory just below the SUSY breaking scale we need now to evolve the techniquark condensates down to the EW scale.  This is achieved by renormalizing the techniquark Yukawa coupling $y_U$ at the SUSY breaking scale, and by simultaneously introducing the dimensionless techniquark renormalization factor
\be
\omega =\frac{\langle U_L \bar{U}_R\rangle_{m_\textrm{SUSY}}}{\langle U_L \bar{U}_R\rangle_{\Lambda_{TC}}}
=\left(\frac{m_\textrm{SUSY}}{\Lambda_{TC}}\right)^{\gamma}\label{omega}\ ,
\ee
 written with  the simple assumption of a constant anomalous dimension $\gamma$ for the techniquark mass operator. 

The four-fermion term involving solely techniquarks is
\be
\frac{y_U^2c_\theta^2}{m_s^2}\omega^2(Q_{Li}\bar{U}_R) (Q_{Li}^* \bar{U}_R^*)
= W_{ijkl}\eta_i^\alpha \eta_{j\alpha} \eta^{*}_{k\beta} \eta^{*\beta}_l\;\;,\;\;
W_{ijkl}=\frac{y_U^2c_\theta^2}{m_s^2}\omega^2
\left(\delta_{ik1}+\delta_{ik2}\right)\delta_{jl3}\ ,
\label{37}
\ee
where $\alpha$ and $\beta$ are spin indices. For this term to be invariant, the spurion $W$ must transform as $W_{ijkl}\rightarrow u_{im}u_{jn}u_{ko}^*u_{lp}^*W_{mnop}$ under $u\in $~SU$(4)$. To estimate the effects of renormalization we made the simplifying factorization assumption leading to an $\omega^2$ effect.

The last term in Eq.\eqref{4fL} derives from decoupling the technisquarks. These come from the Yukawa couplings of the super TC sector (the gauge and the superpotential one) as summarized in the Appendix A.  Decoupling these three complex scalars yields a four techniquark operator respecting the full SU$(4)$ symmetry. This four techniquark operator appears in the Lagrangian as
\be
V \epsilon_{abc}\epsilon_{cde}\eta_i^{\alpha a}\eta_{j\alpha}^{ b} \eta^{* d}_{i\beta} \eta^{*\beta e}_j 
\;\;,\;\;V=\frac{g_{TC}^2}{m_\textrm{SUSY}^2}\omega^2\,, \label{4fermi}
\ee
where we have assumed the same renormalization enhancement factor $\omega$. This operator induces a shift in the coefficient $m^2$ of  the $\rm{Tr} \left[ M M^{\dagger}\right]$ operator. The sign is such that it contributes to chiral symmetry breaking. 

It is also useful to introduce the spurion $\Delta$ related to the soft SUSY breaking mass operator:
\be
\frac{1}{2}M_{D}\bar{D}_{R}\bar{D}_{R}=\eta^{T} \Delta\eta\;,
\ee
with
\be
\Delta=
{\textrm{diag}}\left(0,0,0,\frac{M_D}{2}\right)
\ee 
transforming under SU$(4)$ as $\Delta\rightarrow u^{*} \Delta u^{\dagger}$.

Collecting the information above we write the following new operators emerging at the effective Lagrangian level to the lowest order in the spurions: 
\be
\mathcal{L}_{ETC}=-c_{1}\Lambda_{TC}^2\text{Tr}\left[M \Delta \right]+
c_{2}\Lambda_{TC}^2\text{Tr}\left[MZ\right]+
c_{3}\Lambda_{TC}^4W_{ijkl}M_{ij}M^*_{kl}+ cc.
\label{eeLETC}
\ee
The powers of $\Lambda_{TC}$ are inserted to make the coefficients dimensionless. The coefficients $c_1,c_2,c_3$, parametrize the uncertainty in the couplings of the effective Lagrangian in function of those of the underlying theory. We estimate them by using naive dimensional analysis, and obtain $c_1\sim x^{-1},c_2\sim x^{-1},c_3\sim x^{-2}$, with $x$ defined by
\be
x=\frac{\Lambda_{TC}}{v_w}\ ,
\label{naive}
\ee
and expected to be between $\pi$ and $4 \pi$.

The effective Lagrangian of MSCT, with a strong TC coupling $g_{TC}$ at a scale $\Lambda<\Lambda_{TC}$, is therefore
\be
{\cal L}_{MSCT}^\prime={\cal L}_{SM kin}+{\cal L}_{MWT}+{\cal L}_{ETC}\ ,
\label{FETC}
\ee
with the first term on the right-hand side denoting the covariant kinetic terms of the SM fields, and the other two terms defined respectively in Eqs.(\ref{LMWT},\ref{eeLETC}).

The Lagrangian (\ref{FETC}) depends on the parameters of the MSSM
potential only via the parameters $m_s$ and $t_\theta$ defined in
(\ref{msparameters}). Given the number of tunable parameters in
(\ref{msparameters}), and the fact that the EW symmetry is broken
regardless of the value of $\mu$, it is clear that the $\mu$ parameter
is less constrained than in the MSSM.
To give an example, in the parameter space corresponding to $m_s\approx m_{SUSY}$, Eqs.
(\ref{msparameters}) lead to the following relations:
\be
\mu \approx\frac{\sqrt{2} t_{\theta }}{\sqrt{1-t_{\theta }^2}}m_{\text{SUSY}},\quad b \approx\frac{t_{\theta } \left(1+t_{\theta }^2\right)}{1-t_{\theta}^2} m_{\text{SUSY}}^2\, .
\label{mssus}
\ee

Note that generally $t_\theta<1$ (see Eq. (\ref{sincos})). It will be shown in Section \ref{SPQR} that $t_\theta$ relates to fermion 
masses, and this implies in particular that $t_\theta>0$. 
Hence we see that $\mu$ can be much larger than $m_{SUSY}$ in this slice of the parameter space if
$t_\theta\sim 1$. This slice of the parameter space therefore illustrates that the $\mu$ problem is much weaker in this model 
than in the MSSM. 

Since the Lagrangian (\ref{FETC}) does not depend directly on
$m_{SUSY}$, henceforth we will simply substitute $m_s$ with $m_{SUSY}$.

\section{Spectrum and Precision Tests}
\label{SPQR}
We now investigate the salient aspects of the theory by first analyzing the vacuum properties, then we derive the EW precision test observables, and finally determine the physical spectrum by scanning the theory for the experimentally viable points in the parameter space. 

 \subsection{Electroweak Symmetry Breaking }
\label{EWSBs}
Due to the modification of the low energy effective potential induced by the TC extension, the ground state of theory must be consistently redetermined. We start by searching for a new ground state parameterized by the following vacuum matrix:  
 \be
\langle M\rangle =\frac{1}{\sqrt{2}}\left(
\begin{array}{cccc}
 0 & 0 & v_1 & 0 \\
 0 & \sqrt{2} v_2 & 0 & v_3 \\
 v_1 & 0 & 0 & 0 \\
 0 & v_3 & 0 & \sqrt{2} v_4
\end{array}
\right)\ .
\label{Mvev}
\ee
This vacuum ansatz leads to the correct EW symmetry breaking pattern. By minimizing the scalar potential given in Eq.(\ref{FETC})  we get the conditions for the minimum:
\bea
c_1 \Lambda_{TC}^2 M_D&=&2 \left(v_2^2-v_4^2\right) \left(2 v_4 \lambda '-\frac{v_1^2 \lambda ''}{v_2}\right),\ c_3 \Lambda_{TC}^4\frac{y_U^2 c_\theta^2}{m_\textrm{SUSY}^2}\omega^2=2 \left(v_1^2-2 v_2^2\right) \left(\lambda '-\frac{v_4 \lambda
   ''}{v_2}\right),\nonumber \\
   v_3&=&0,\ m^2=\lambda  \left(v_1^2+v_2^2+v_4^2\right)+4 \lambda ' v_2^2+\frac{2 v_1^2 v_4 \lambda ''}{v_2} .
\label{vevc}
\eea
Within this ansatz the minimum is a global one. Note that this solution is not smoothly connected to the MWT one since that appears only when $M_D=y_U =0$. The scalar physical spectrum in the unitary gauge consists of: 4 neutral scalars ($h_1, h_2, h_3, \phi$), 3 neutral pseudoscalars ($A_1, A_2, A_{\phi}$), 3 charged scalars ($h_1^{\pm}, h_2^{\pm},h_3^{\pm}$) and 2 doubly charged scalars $(h_1^{\pm\pm},h_2^{\pm\pm})$. The mass spectrum is computed in Appendix \ref{AppA}. 
The masses of the upper component $u$ and the lower one $d$ of a generic EW fermion doublet are given by
\be
m_u=c_2 \Lambda_{TC}^2\frac{c_{\theta }^2 y_U y_u \omega}{m_\textrm{SUSY}^2}\frac{v_1}{\sqrt{2}} ,\qquad  \qquad m_d=\frac{y_d}{y_u}t_{\theta} m_u\, .
\label{mfglobal}
\ee 
From the last equation and the LEP lower bound on the chargino masses, $m_E,m_N\gtrsim 100$ GeV, it follows $t_{\theta}>0$,
and we have that in general $t_{\theta}<1$. 

Having established the tree-level vacuum properties of the model we can now turn to the phenomenological consequences of the model. In the next two subsections we investigate different values of $m_\textrm{SUSY}$ and Yukawa couplings yielding the correct SM mass spectrum. 

\subsection{Precision Tests}

We now determine the EW precision parameters \cite{Peskin:1990zt,Peskin:1991sw,Kennedy:1990ib,Altarelli:1990zd} which receive contributions coming from the vacuum structure above, the intrinsic TC dynamics as well as the new lepton sector. 
It is the combination of these contributions which lead to a phenomenological viable model. 

\subsubsection{Tree-level contribution}
To this order order the main effect on precision observables comes from the vacuum properties of the model. For the VEV defined by Eqs.(\ref{Mvev},\ref{vevc}) the $W^\pm$ and $Z$ bosons squared masses are readily computed and read:  
\be
m_Z^2=\frac{1}{4} \left(g_L^2+g_Y^2\right) \left(v_1^2+4 v_2^2\right),\ m_{W^\pm}^2=\frac{1}{4} g_L^2 \left(v_1^2+2 v_2^2\right)\ .
\ee
From these expressions one determines the value of the $T$ parameter at tree level:
\be
\alpha_e T_{tree}=-\frac{2 v_2^2}{v_w^2}, \qquad\ v_w^2=(\sqrt{2}G_F)^{-1}=v_1^2+2 v_2^2=\left(246\ {\rm GeV}\right)^2\ ,
\label{tvw}
\ee
where $G_F$ is the Fermi coupling constant and $\alpha_e$ the fine-structure constant. We note that extensions of the SM can have a tree-level nonzero $T$ parameter while remaining phenomenologically viable \cite{DiChiara:2008rg}. The tree-level value of the $S$ parameter vanishes.

\subsubsection{Beyond tree level}
At one loop there are two distinct contributions to both $S$ and $T$, one coming from the $N$ and $E$ heavy leptons with non-degenerate masses \cite{He:2001tp}, and the other one generated by the techniquarks $U^a$ and $D^a$. The latter contribution is usually accounted for by the naive values $S_{naive}=(6\pi)^{-1},\ T_{naive}=0$, associated to each EW doublet, that are obtained in the limit of degenerate $U^a$ and $D^a$ having a dynamically generated mass $m_{dyn}\rightarrow\infty$. In the present article we estimate $S$ and $T$ by taking into account the mass splitting of  $U^a$ and $D^a$, as well as the finiteness of $m_{dyn}$. We use the formulas and notation for the contributions of Majorana mass terms to EW parameters as given in  \cite{Frandsen:2009fs}, with the mass parameters multiplying the operators $U_L\bar{U}_R,\ D_L\bar{D}_R,\ D_R D_R$   given by
\be
\frac{y_U^2\omega^2 c_\theta^2}{m_\textrm{SUSY}^2} \langle U_L\bar{U}_R\rangle_{\Lambda_{TC}}+m_{dyn}\ , \quad m_{dyn}\ ,\quad M_D,
\label{mMj}
\ee
determined using Eq.(\ref{4fL}). These masses are identified respectively with $m_\zeta$, $m_D$, and $m_R$ of \cite{Frandsen:2009fs}. The expressions above are dictated by dimensional analysis and symmetries. Similar estimates, albeit without any knowledge of the ETC dynamics,  have been used in the literature \cite{Hill:2002ap}.
We assume as a simple estimate for the dynamical quantities above 
\be
m_{dyn}\simeq\frac{x \, v_w}{2}\ ,\qquad \langle U_L\bar{U}_R\rangle_{\Lambda_{TC}} \simeq x v_1^3 \ ,
\ee
with $x$ given in Eq.\eqref{naive}, providing the link between $\Lambda_{TC}$ and the electroweak scale. 

\begin{figure}[h]
\begin{center}
\includegraphics[width=12pc]{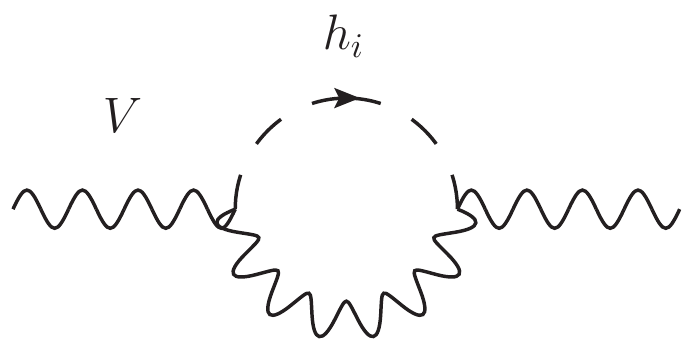}
\caption{\label{fig:HtoST} Higgs scalar $h_i$ contribution to vector boson V's vacuum polarization amplitude.}
\vspace{1pc}
\end{center}
\end{figure}

The experimental bounds on $S$ and $T$ depend on the reference mass of the SM Higgs, $m_{h_{SM}}$. As it is shown in the next section, vacuum stability imposes the constraint $v^2_1>v_2^2$, and moreover the large value of $m_t$, which is proportional to $v_1$, favors $v_1\gg v_2$. Therefore to determine the corresponding contributions in MSCT we work in the limit $v_2/v_1\rightarrow 0$, which further simplifies the calculation. In this limit the gauge eigenstate $\sigma_U^0$, which has the same couplings to the EW gauge bosons of the SM Higgs, is a linear combination of just two mass eigenstates
\be
\sigma_U^0=c^\prime_j h_j,\quad i=1,2,
\ee
with the sum of the squares of the coefficients $c^\prime_j$ normalized to one.
The only nonzero contributions of $h_i$ to $S$ and $T$ are associated with the diagrams of the kind given in Fig.\eqref{fig:HtoST}. 
In the limit of $m_{h_i}\gg m_Z$ the resulting expressions for $S$ and $T$ in MSCT corresponding to the SM Higgs contributions are:
\be
S\approx\frac{1}{12\pi}\sum_i c_i^{\prime 2} \log\frac{m_{h_i}^2}{m^2_{Ref}}\ ,\quad T\approx\frac{-3}{16\pi c_w^2}\sum_i c_i^{\prime 2} \log\frac{m_{h_i}^2}{m^2_{Ref}},
\ee
with $c_w$ the cosine of the Weinberg angle, and $m_{Ref}$ conventionally defined equal to 117 GeV. The expressions above reduce to the $h_{SM}$ result if one takes
\be
m_{h_{SM}}=\prod_{i=1}^2m_{h_i}^{c_i^{\prime 2}} \ .
\label{mHg}
\ee
The numerical value we use for the reference Higgs mass when comparing the MSCT predictions to the experimental limits on $S$ and $T$ is given by Eq.\eqref{mHg}.  

To summarize, the full contribution to the EW parameters is given by
\be
S=S_{U,D}+S_{N,E}\ ,\qquad T=T_{tree}+T_{U,D}+T_{N,E} \ ,
\ee
where $T_{tree}$ is the tree-level contribution of Eq.(\ref{tvw}), $S_{U,D}$ and $T_{U,D}$ are the one loop contributions of the techniquarks calculated by using the formulae given in \cite{Frandsen:2009fs}, and $S_{N,E}$ and $T_{N,E}$ are the one loop contributions of the heavy leptons \cite{He:2001tp}.

\subsection{Viable Mass Spectrum}
\label{VMSs}
In this section we show that MSCT is phenomenologically viable in a large region of its parameter space, and we determine general features of the mass spectrum.

We first define a useful parameter:
\be
m_r^2\equiv 2 \left(\frac{v_4}{v_2}-\frac{v_2}{v_4}\right) \left(2 v_2 v_4 \lambda '-v_1^2 \lambda ''\right).
\ee
We impose constraints on the parameter space by requiring the extremum of the potential to be a global minimum. To do that we start by requiring the squared masses $m_{\phi}^2,m^2_{A_1},m^2_{h^\pm_1},m^2_{h^{\pm\pm}_1}$, given in Eqs.(\ref{mh4},\ref{mpi},\ref{mhp},\ref{mhpp}), to be positive. This, together with the definition of the EW VEV, Eq.(\ref{tvw}), determines the bounds
\be
v_w\geq v_1>\frac{v_w}{\sqrt{2}},\  m_r^2>0,\ m_{h^{\pm\pm}_1}>0,\ \left|v_4\right|>\sqrt{\frac{v_w^2-v_1^2}{2}}.
\ee
Notice that in the previous constraints we have chosen $v_1>0$. 

Expecting a dynamical mechanism in the TC sector we take $m^2 > 0$ in Eq.\eqref{su4}. We furthermore require the potential to be stable by imposing Eqs.\eqref{stab}.
We have scanned the parameter space in the region defined by
\bea
&&0.5<c_{1,2} x<5,\  0.5<c_3 x^2<5,\   \gamma=1.5,\ 246\ \geqslant v_1\ {\rm GeV}^{-1} \geqslant  246/\sqrt{2},\ \nonumber \\
&&\left(2\pi\right)^2>\lambda>0.1,\ 2\pi>y_{t,N,E,U}>0.1,\ 4\pi>g>\pi,\ m_t=172\ {\rm GeV},\nonumber \\
&& 120\ {\rm GeV}<m_r<10\ {\rm TeV},\ \left|v_4\right|>\sqrt{\frac{v_w^2-v_1^2}{2}}\ ,\  m_{\rm SUSY}>5\ {\rm TeV}.
\label{scanR}
\eea
We have allowed the Yukawa region to extend to large values since we argued in \cite{Antola:2010jk} for the possible presence of a UV fixed point around $2 \pi$. Of course, this estimate must be taken with a grain of salt given that nonperturbative corrections will also affect the results. 

Because of the four techniquark operators  in Eqs.~(\ref{37},\ref{4fermi})  and based on \cite{Fukano:2010yv} it is reasonable to expect the anomalous dimension of the techniquark condensate to be larger than unity but smaller than two. For definiteness we have  assigned to it the value of 1.5. Smaller values of $\gamma$ are also phenomenologically viable and lead, in general, to a lighter spectrum.  

The region of the parameter space we explored is inspired by a naive dimensional analysis estimate for the expected values of the couplings. We defined a point in parameter space as viable if: The point corresponds to a global minimum of the potential; All the scalars have masses greater than 120 GeV; The heavy leptons $N$ and $E$ have masses greater than 100 GeV; The experimental bounds on $S$ and $T$ at 90\% CL are satisfied. We moreover imposed $m_\textrm{SUSY}> 5$ TeV, to implement the relation $m_\textrm{SUSY}>\Lambda_{TC}$ used in the derivation of the effective Lagrangian, Eq.\eqref{eLETC}. Because in the effective Lagrangian approach the techniquarks are required to be lighter than the cutoff scale, we imposed the Dirac and Majorana masses, defined in Eqs.(\ref{mMj}), to be smaller than $\Lambda_{TC}$. 

The time needed to complete the scan  of the parameters space grows with decreasing $v_1$, since the viable points became more rare in this region of the parameter space, and so we chose arbitrarily to limit $v_1$ to values larger than 210 GeV, and stopped the scan once we collected 10000 viable points.

\begin{figure}[h]
\begin{center}
\includegraphics[width=20pc]{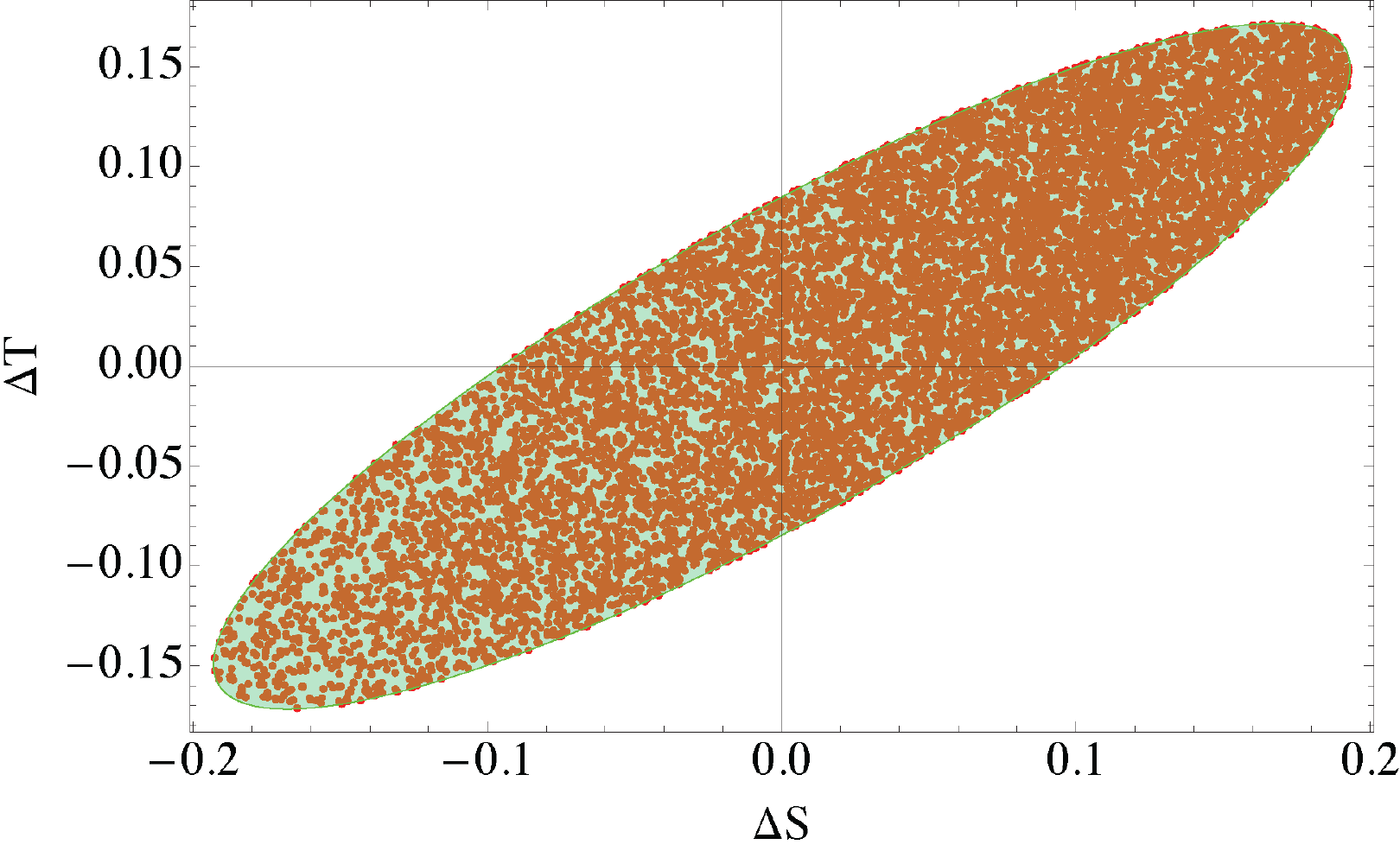}
\caption{\label{fig:ST}The axes are defined as the respective difference between $S$ and $S_{min}$, on the $x$ axis, and  $T$ and $T_{min}$, on the $y$ axis. The shaded area delimited by the ellipse represents the experimentally viable region at 90\% CL.  The orange points represent each a set of parameters for which MSCT is viable.}
\vspace{1pc}
\end{center}
\end{figure}

 In Fig.\eqref{fig:ST} we give the difference between the MSCT prediction on the $S$ and $T$ parameters and the corresponding experimental limits, which depend on the reference Higgs mass, that we defined in MSCT by Eq.\eqref{mHg}. As required all the 10000 viable points lie within the 90\%CL ellipse, with about 63\% of them being in the $\Delta S>0$ region. This is typical for TC theories, for which degenerate techniquarks give a net positive contribution to $S$. In MSCT the positive contribution of $S_{U,D}$ is reduced by the mass splitting of $U$ and $D$ induced by the ETC interactions, and partially canceled by $S_{N,E}$, which is negative for $m_N<m_E$.

\begin{figure}[h]
\begin{center}
\includegraphics[height=13pc]{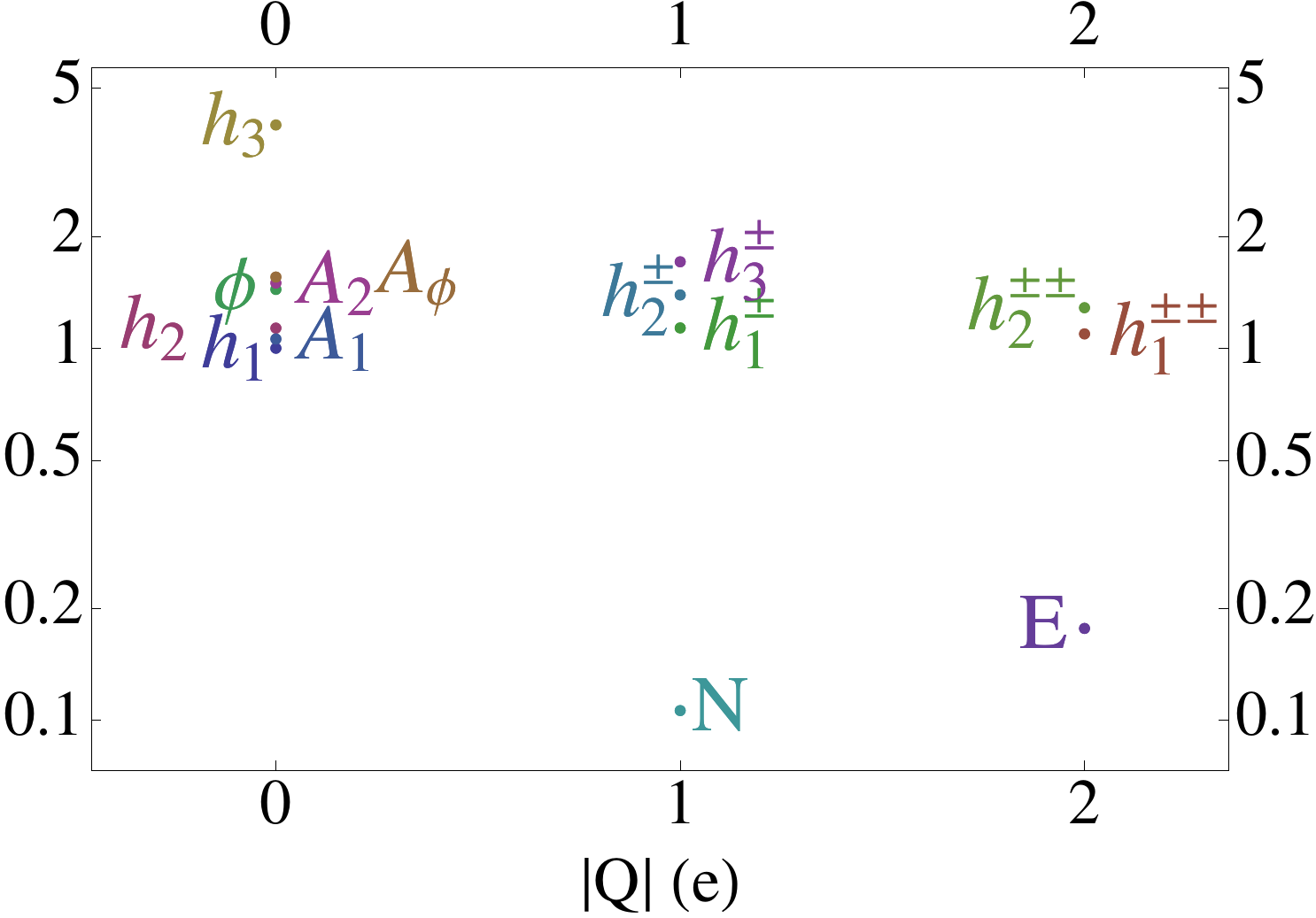}
\includegraphics[height=13pc]{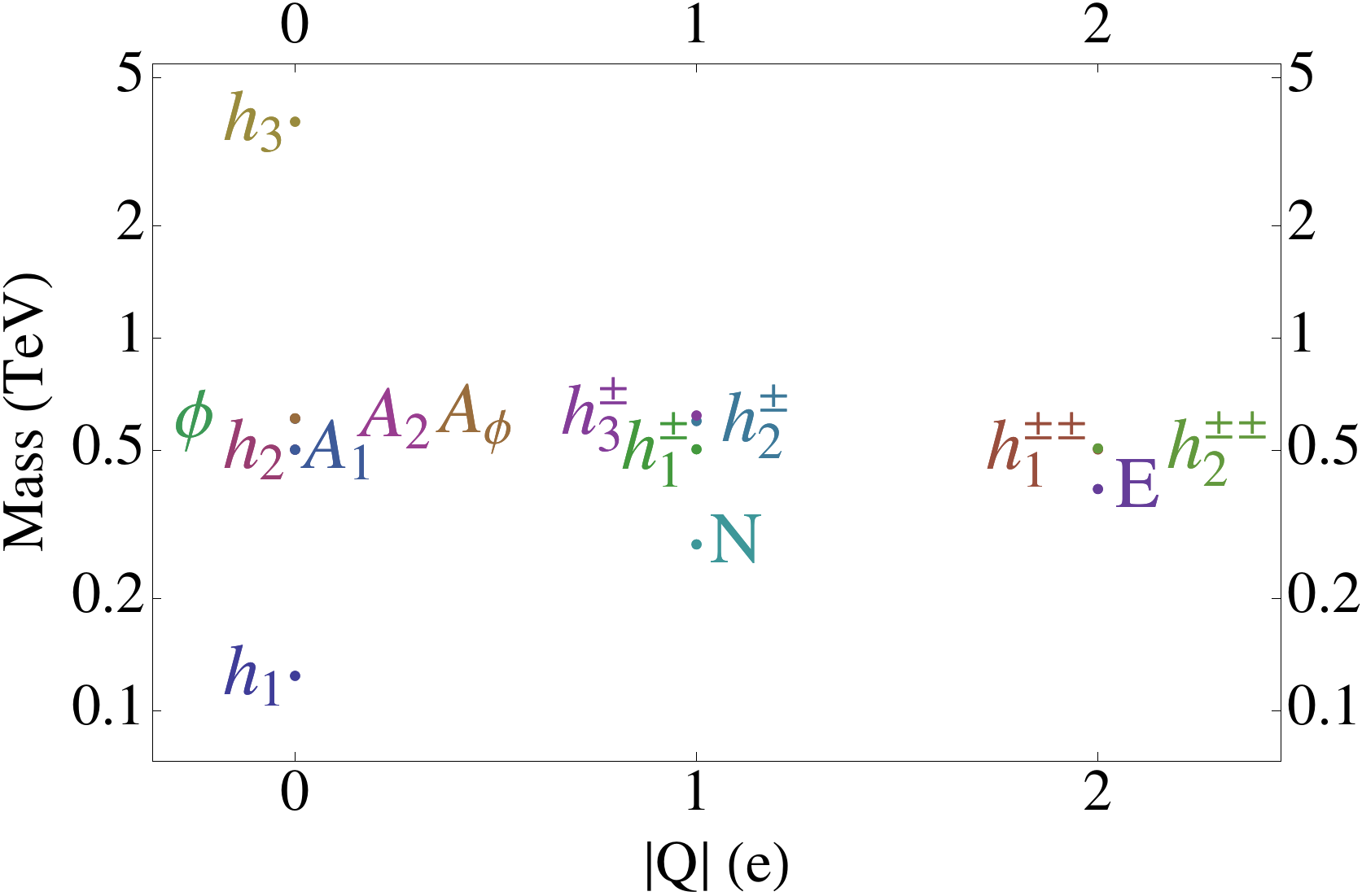}
\caption{\label{fig:heavylightH}Scalar and heavy lepton mass spectrum in function of the EM charge Q corresponding to a heavy (light) composite Higgs, with $m_{h_1}\simeq 1$ TeV (left panel) and $m_{h_1}\simeq 125$ GeV (right panel).}
\vspace{1pc}
\end{center}
\end{figure}

In Fig.\eqref{fig:heavylightH} we present, for illustrative purposes, the mass spectra for the MSCT non-SM particles associated to two of the 10000 data points plotted in Fig.\eqref{fig:ST} , corresponding respectively to a heavy (1 TeV) and light (125 GeV) lightest neutral Higgs scalar, with the absolute value of the EM charge on the $x$ axis. These two particular data points reflect general correlations in the mass spectrum that we investigate in detail in the rest of this section. The scalars and pseudoscalars mass eigenstates are a mixture of the  original MWT  spin zero composite states. From Fig.\eqref{fig:heavylightH} it is already clear that the scalars and pseudoscalars are generally heavier than the leptons. From these figures we can also infer that, as discussed earlier in Sec. \ref{effMWT} (see the discussion around Eq. (\ref{mwtgoldstonemass})), there are no unwanted light pseudo-Goldstone states in the physical spectrum.

\begin{figure}[h]
\begin{center}
\includegraphics[width=19pc]{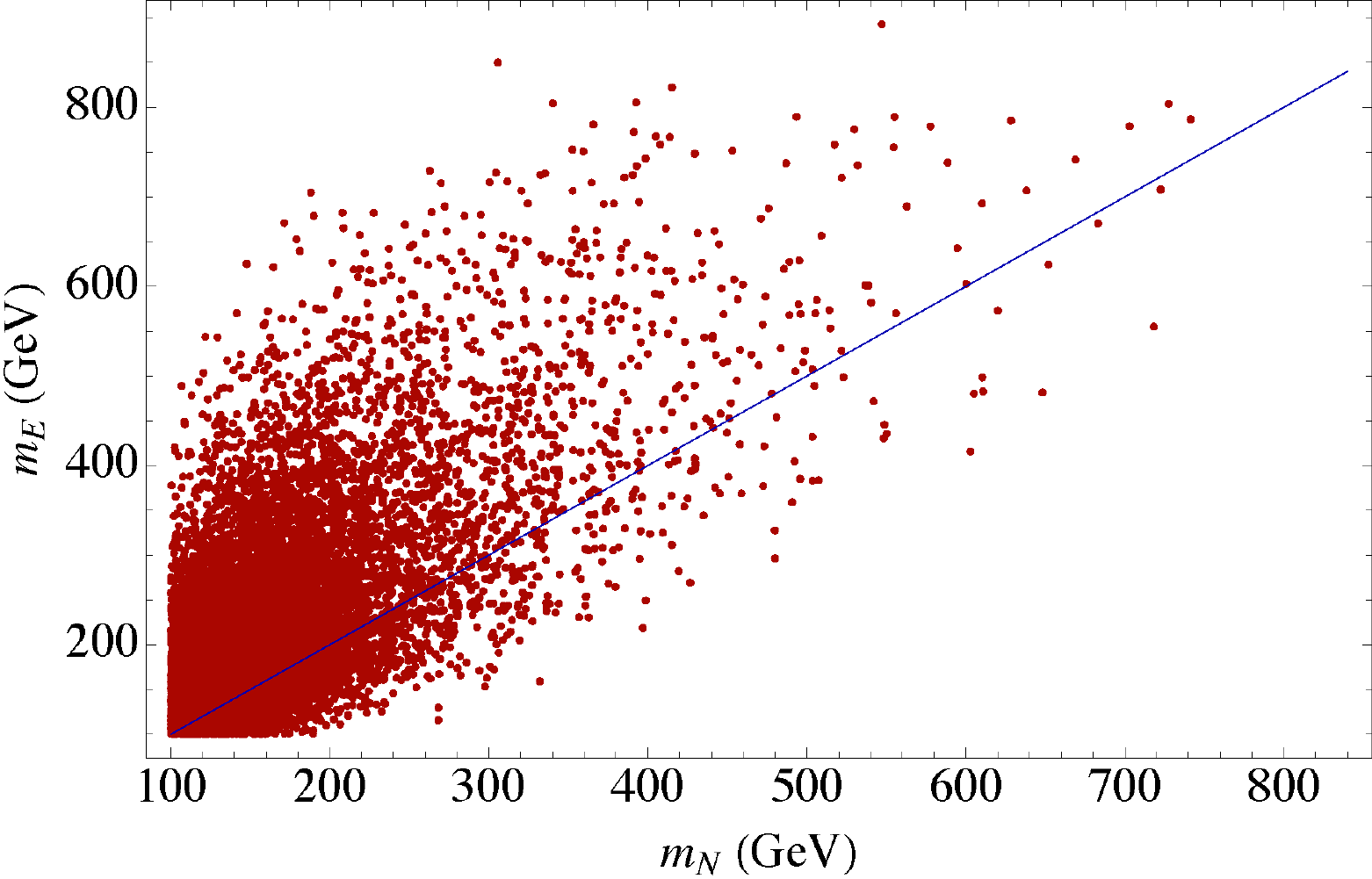}
\includegraphics[width=19pc]{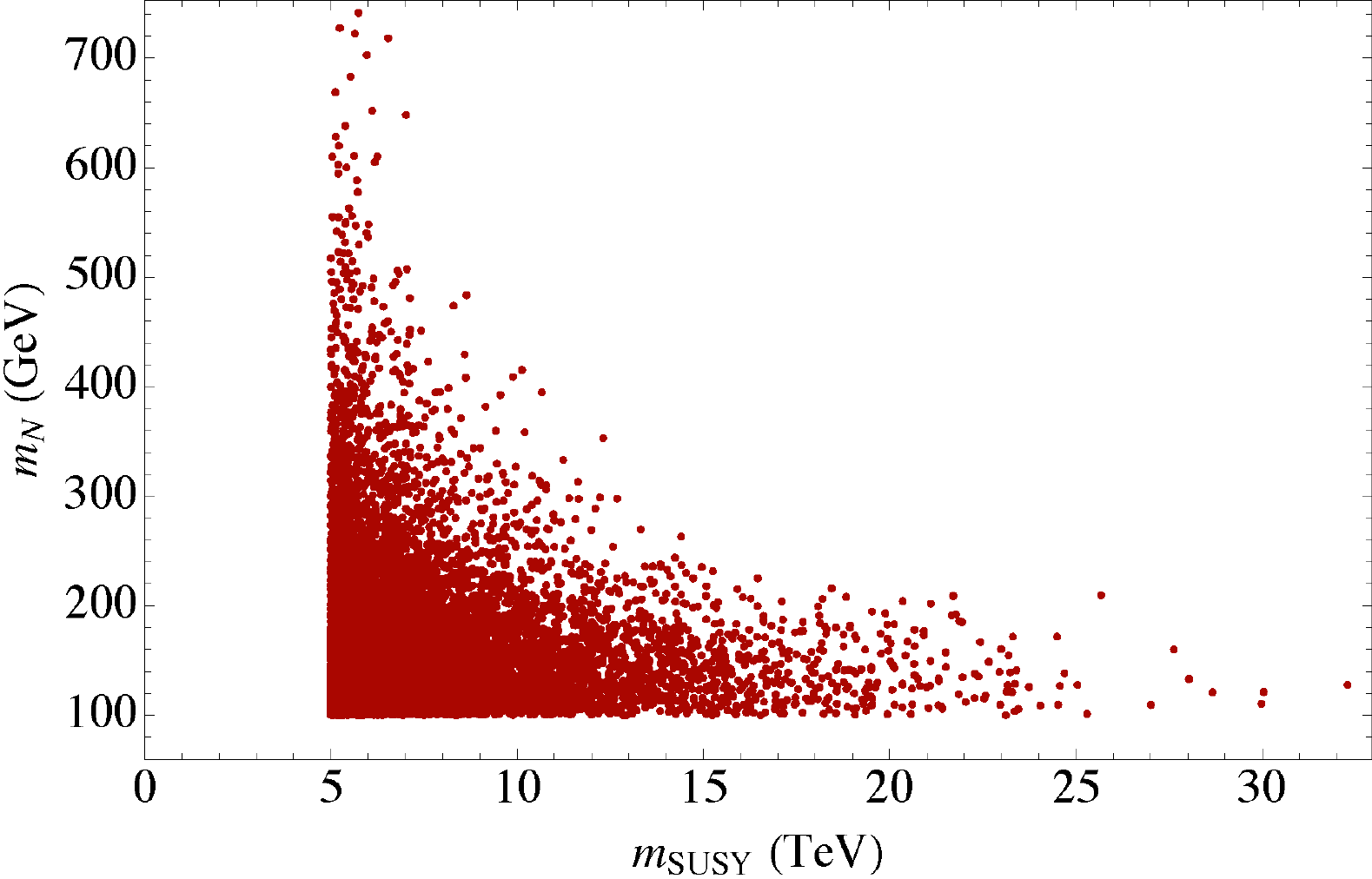}
\caption{\label{fig:mNmE} On the left panel the mass of $E$ versus that of $N$ is plotted. The blue straight line lies at $m_N=m_E$. On the right panel the mass of the heavy lepton $N$ vs the SUSY breaking scale $m_\textrm{SUSY}$ is shown.}
\vspace{1pc}
\end{center}
\end{figure}

The mass of the heavy leptons $N$ and $E$ is of $O(v_w)$, as it can be seen in Fig.\eqref{fig:mNmE}, left panel. This result is determined by the fact that the new leptons receive masses via Yukawa interactions which we required to be smaller than $4\pi$. For most of the viable points $m_N<m_E$, which yields $T_{N,E}>0$: indeed a positive one loop contribution to $T$ is necessary to offset the large negative tree-level contribution, $T_{tree}$. In the right panel of Fig.\eqref{fig:mNmE} the dependence of $m_N$ on the SUSY breaking scale is shown. The mass of the fermions is inversely proportional to $m_\textrm{SUSY}^{2-\gamma}$, which for $\gamma=1.5$ gives a maximum of $m_N$ following approximately this function. An interesting result is the rather low mean value of the SUSY breaking scale, around 8 TeV, with less than 1\% of the viable points featuring $m_\textrm{SUSY}>19$ TeV. 

Most viable points satisfy $m_N \gtrsim m_E$. The heavy lepton number associated to $N$ and $E$ is conserved in the approximation that we have made by neglecting the contribution of the coupling $y_R$ in Eq.\eqref{spmwt}, and therefore the lighter of the two charginos is stable. Even including the contribution in the effective Lagrangian of terms proportional to $y_R$, in the most common case of $m_N<m_E$ the chargino $N$ is stable. Unfortunately a stable charged dark matter candidate is basically ruled out at the TeV scale \cite{Nakamura:2010zzi}. This problem can be easily circumvented by adding lepton number violating interactions in the superpotential allowing $N$ to decay to leptons of the third family. 

Another variation of  MSCT free from doubly charged leptons can also be easily constructed by adding three SM like heavy lepton families rather than the one with exotic electric charges. We performed a parameter space scan also for this model and found it to be phenomenologically viable. The associated mass spectrum is very similar to that of MSCT in the same limit $m_\textrm{SUSY}>\Lambda_{TC}$.  

\begin{figure}[h]
\begin{center}
\includegraphics[width=19pc]{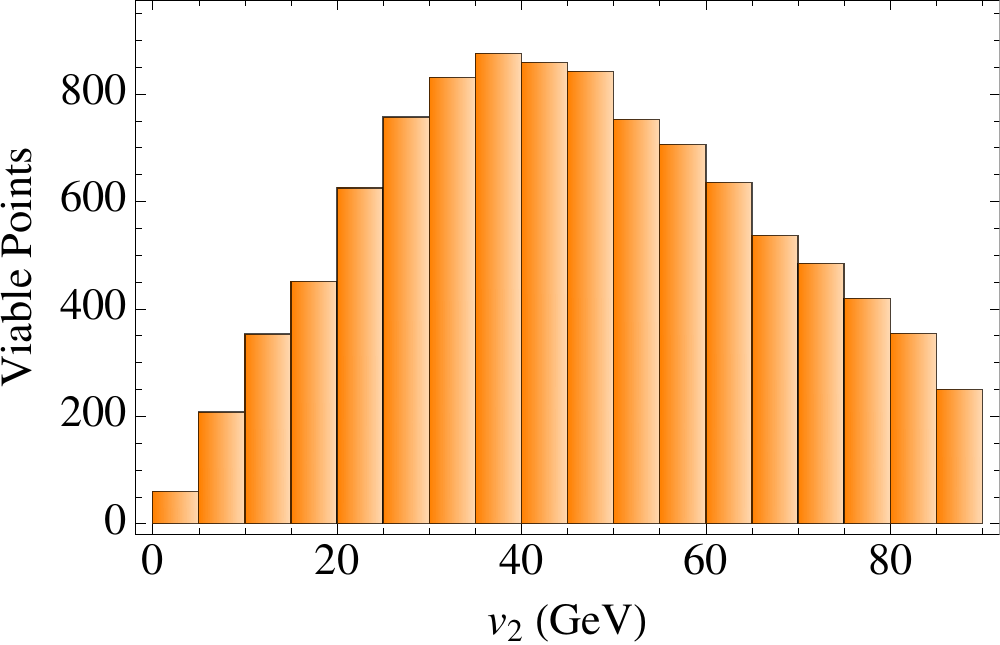}
\includegraphics[width=18.5pc]{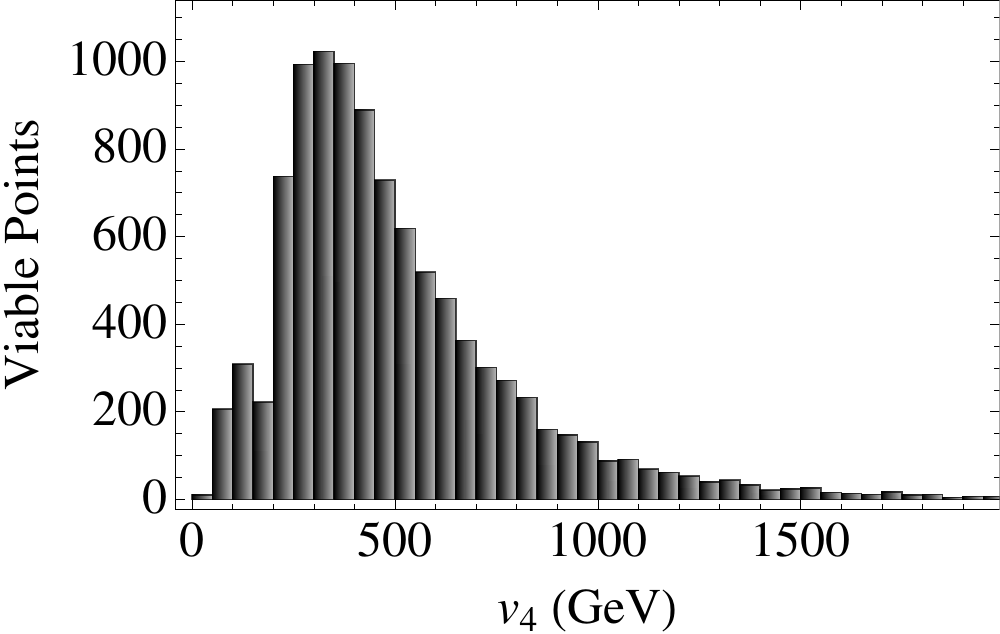}
\caption{\label{fig:v24N} On the left (right) panel is shown the histogram with the number of viable point in function of $v_2$ ($v_4$).}
\vspace{1pc}
\end{center}
\end{figure}

The heavy scalar mass spectrum is generated mostly through ETC interactions that induce nonzero but small values of $v_2$, as seen in the left panel of Fig.\eqref{fig:v24N}, with a maximum of the distribution at approximately $38$ GeV. Large values of $v_4$ (Fig.\eqref{fig:v24N}, right panel) are required to generate a Majorana mass for $D_R$ nearly degenerate with the mass of $U$, which makes $T_{U,D}$ small and helps to bring the $T$ parameter into agreement with the experimental constraints. A comment on vacuum alignment  is in order. 
We have not determined here the quantum corrections which can provide a partial vacuum re-alignment \cite{Peskin:1980gc,Preskill:1980mz} in the MWT vacuum direction \cite{Dietrich:2009ix}. 

\begin{figure}[h]
\begin{center}
\includegraphics[width=20pc]{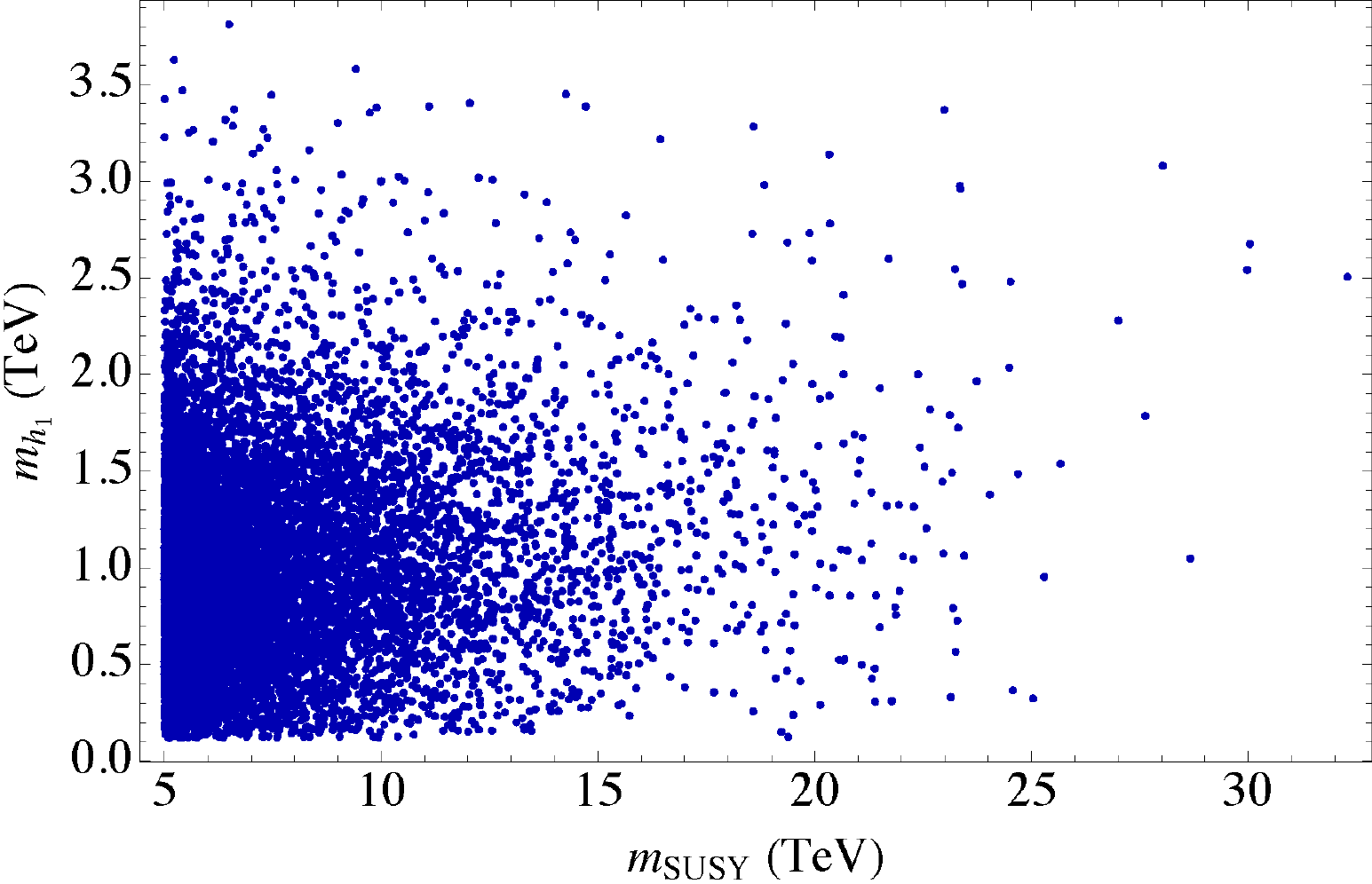}
\caption{\label{fig:msusymh} Dependency of the mass of the lightest neutral scalar, $m_{h_1}$, on the SUSY breaking scale.}
\vspace{1pc}
\end{center}
\end{figure}

\begin{figure}[h]
\begin{center}
\includegraphics[width=18.9pc]{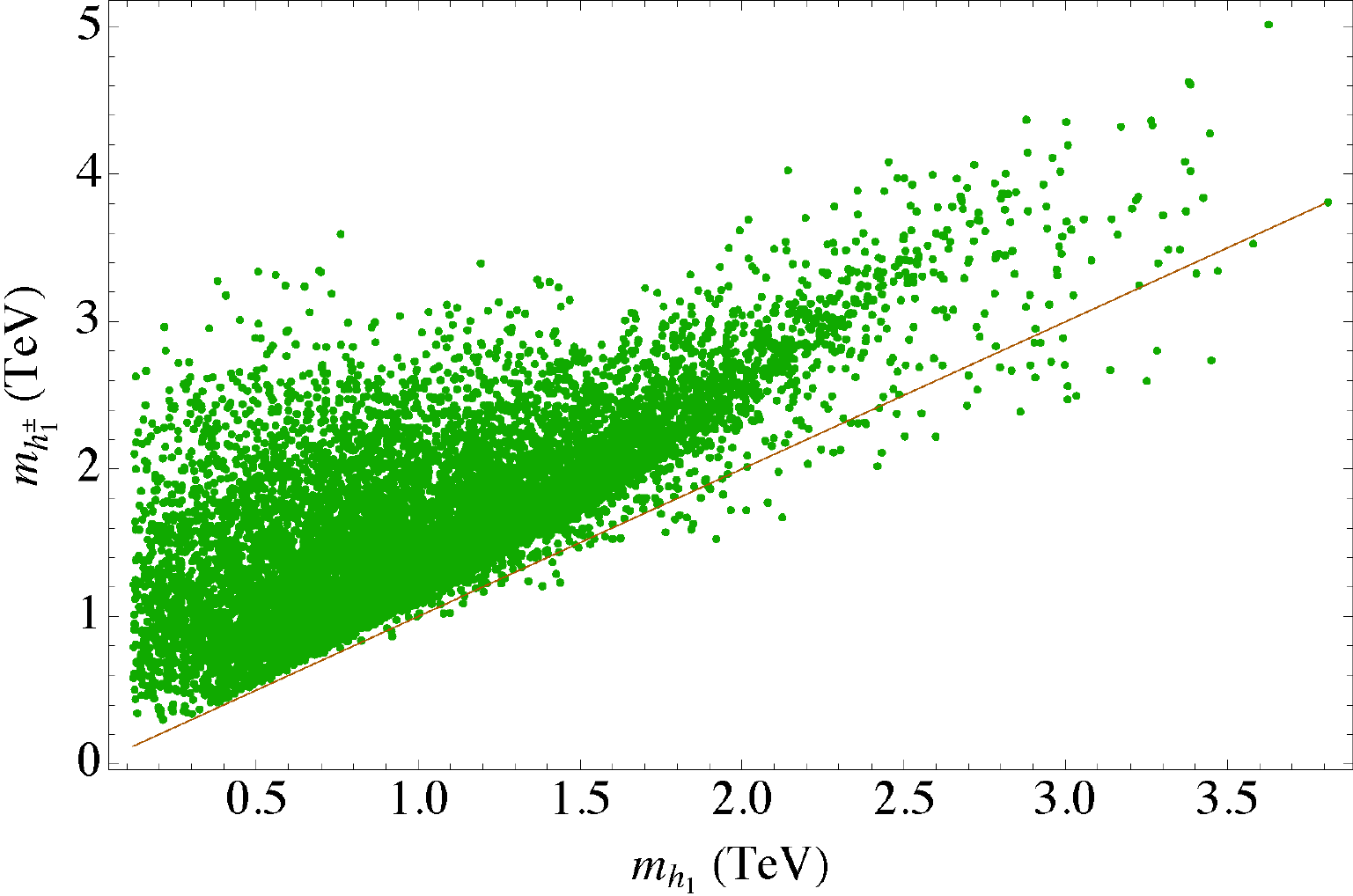}
\includegraphics[width=19.3pc]{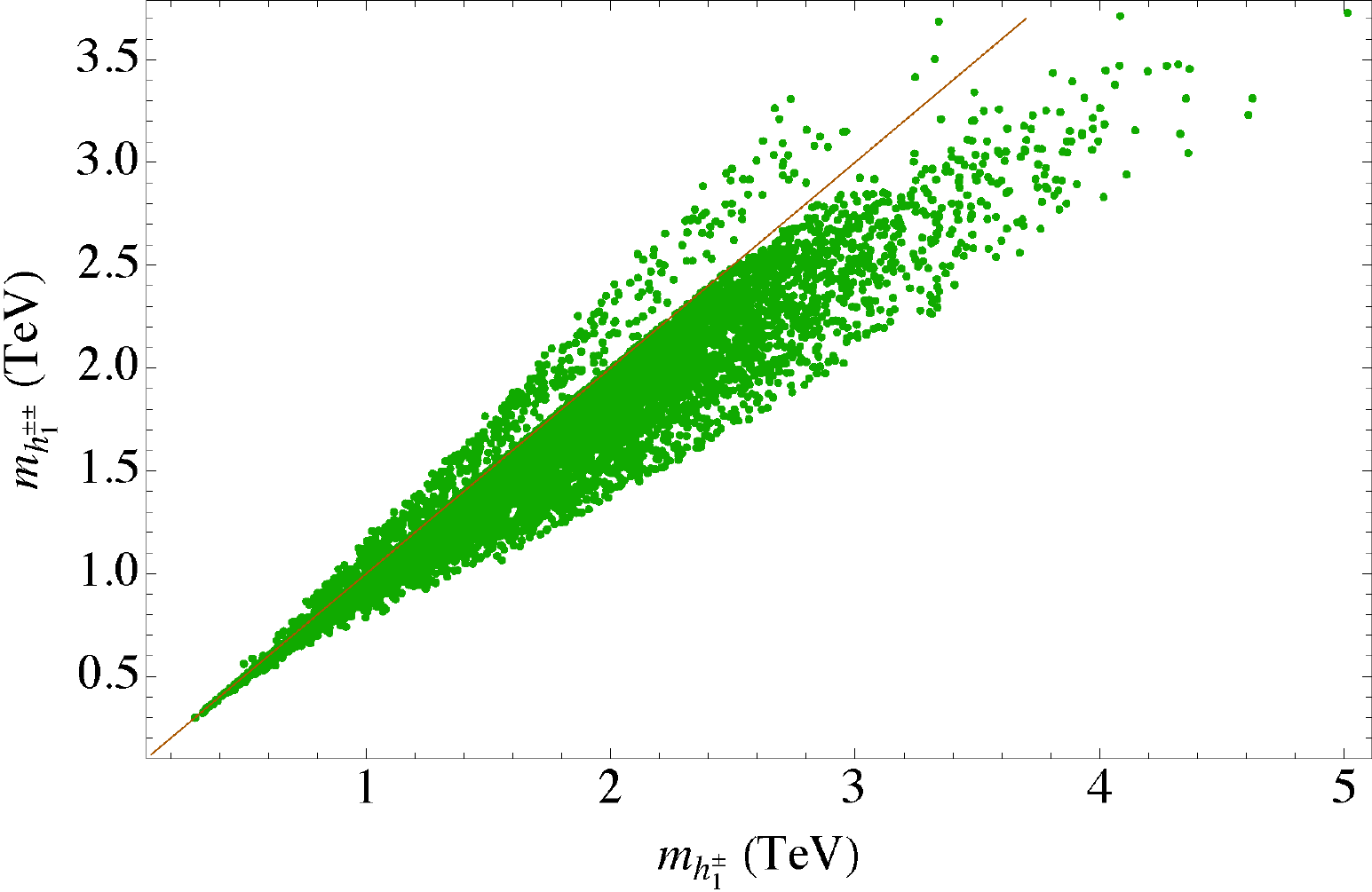}
\caption{\label{fig:mhmhp} On the left (right) panel is plotted the mass of the lightest charged (doubly charged) scalar in function of the mass of the lightest neutral (charged) Higgs scalar. The orange straight line lies at $m_{h^\pm_1}=m_{h_1}$ ($m_{h^\pm_1}=m_{h^{\pm\pm}_1}$).}
\vspace{1pc}
\end{center}
\end{figure}

The mass of the lightest Higgs scalar does not show any dependence on the SUSY breaking scale, and is in general rather heavy, as it can be seen in Fig.\eqref{fig:msusymh}, with less than 1\% of the viable points featuring $m_{h_1}<2 m_W\cong 160$ GeV and an average mass equal to about a TeV. This is indeed what naive dimensional analysis suggests, but it is still an interesting result since we allowed the couplings in the effective Lagrangian to take also rather small values associated with nearly conformal (or walking) dynamics.
A clear mass hierarchy is manifest in Fig.\eqref{fig:mhmhp} for the neutral and charged lightest scalars, with $m_{h^\pm_1}>m_{h_1}$ (left panel) and $m_{h^{\pm}_1}\gtrsim m_{h_1^{\pm\pm}}$ (right panel) for almost the whole sample of viable points.

In the next section we will study the potential of LHC to produce and observe one of the composite neutral Higgs scalars predicted by MSCT. In particular we will focus our attention on the neutral scalar with the strongest linear coupling to $W$, which we label simply by $H$: this scalar corresponds in most cases to either $h_1$ or $h_2$, or otherwise to $h_3$.

\section{Higgs Phenomenology at the LHC}
\label{LHCPh}

The sector of the MSCT effective Lagrangian relevant for Higgs production and decay at LHC is given by the linear Higgs coupling terms
\bea
{\cal L}_{int}&=&-2 m_{h^\pm}^2 h^+ h^-\frac{\lambda_i}{v_w} h_i-m_f \psi \bar{\psi} \frac{c_j}{v_1} h_j+2 m_W^2
 W^+_\mu W^{\mu -} \left(\frac{v_1}{v_w^2} c_j h_j+\frac{2 v_2}{v_w^2} d_j h_j\right)\nonumber \\
&+&m_Z^2
Z_\mu Z^{\mu} \left(\frac{v_1}{v_w^2+2 v_2^2} c_j h_j+\frac{4 v_2}{v_w^2+2 v_2^2} d_j h_j\right),
\label{HlinL}
 \eea
where we suppressed indices and sums over the MSCT fermions and charged composite scalars, represented in the equation above by $\psi$ and $h^\pm$, and the sum over $j=1,2,3$. The coefficients $c_j,\ d_j$ are defined by
\be
\sigma_U^0=c_j h_j,\quad \Delta^0=d_j h_j,
\ee
and are obtained by inverting the mixing matrix $U_s$ in Eq.\eqref{mixM}. We calculated numerically $c_j, d_j$, and $\lambda_j$ for each viable point by expressing the Lagrangian in terms of the mass eigenstates and reading off the relevant couplings.

\begin{figure}[h]
\begin{center}
\includegraphics[width=28pc]{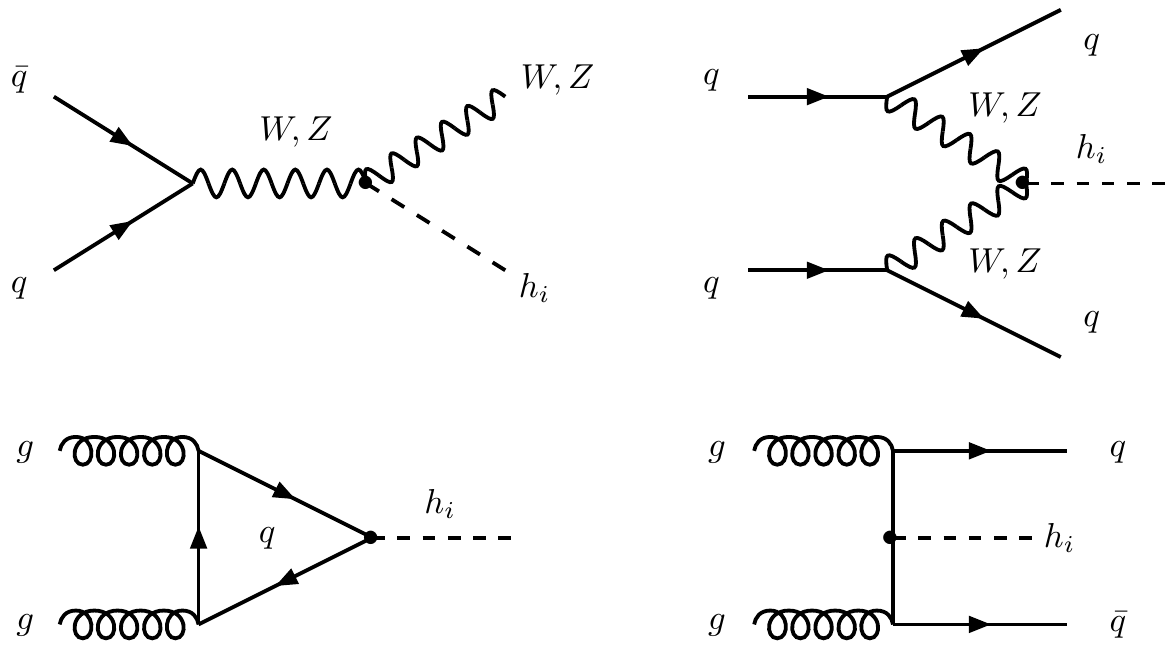}
\caption{\label{fig:Hprod} Dominant neutral Higgs scalar $h_i$ production processes at LHC.}
\vspace{1pc}
\end{center}
\end{figure}

\begin{figure}[h]
\begin{center}
\includegraphics[width=28pc]{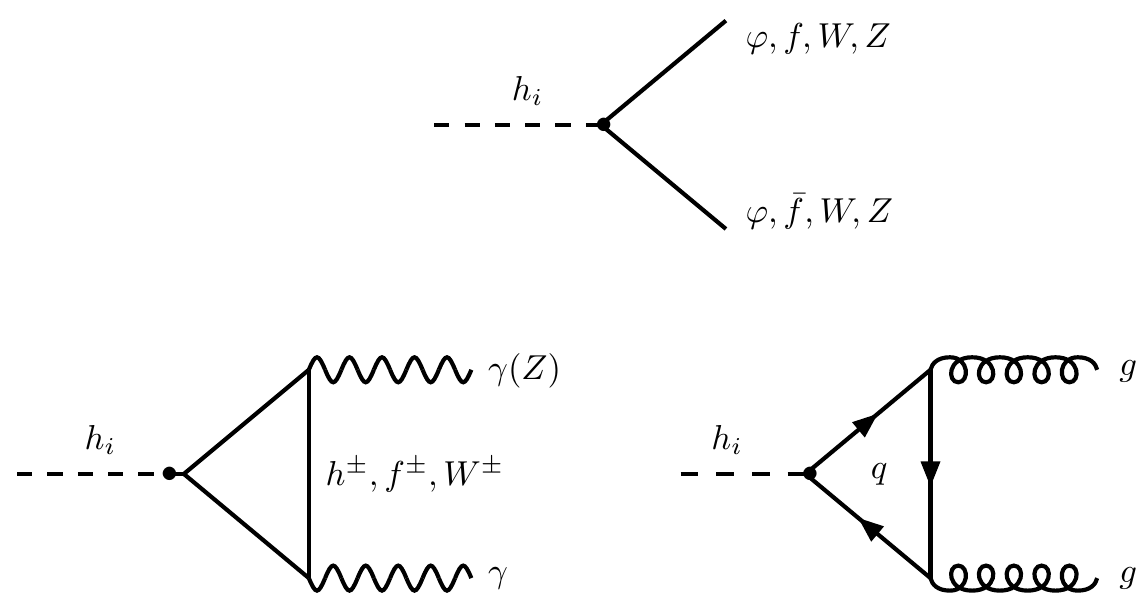}
\caption{\label{fig:Hdec} Decay channels for the neutral Higgs scalar $h_i$.}
\vspace{1pc}
\end{center}
\end{figure}

The main production processes for a Higgs neutral scalar at LHC (see \cite{Gunion:1989we} by Gunion et al., and \cite{Djouadi:2005gi} by Djouadi for comprehensive reviews on Higgs phenomenology at LHC) are $q\bar{q} \longrightarrow W/Z+ H$, $q q \longrightarrow qq+ H$, $g g  \longrightarrow H$, $g g\longrightarrow q\bar{q}+H$, with the corresponding Feynman diagrams shown in the same order in Fig.\eqref{fig:Hprod}.
The Higgs scalar will then decay to the whole spectrum of massive particles and, through loop induced processes involving charged (colored) massive particles, to photons (gluons) as well, as shown in Fig.\eqref{fig:Hdec}, where $\varphi$ represents any spin 0 particle.
From Figs.(\ref{fig:Hprod},\ref{fig:Hdec}), Eq.\eqref{HlinL}, and the fact that the fermion couplings to gauge bosons are fixed by their quantum numbers, it follows that the dominant production rates as well as the decay rates to fermions, $W/Z$ bosons (both on and off shell), and gluons, can be expressed in function of the corresponding SM rates by:
\bea
\frac{\sigma_{q\bar{q}\rightarrow W h_i}}{\sigma_{q\bar{q}\rightarrow W h_{SM}}}&=&\frac{\sigma_{q q\rightarrow q^ \prime q^ \prime h_i}}{\sigma_{q q\rightarrow q^\prime q^\prime h_{SM}}}=\frac{\Gamma_{h_i\rightarrow W W}}{\Gamma_{h_{SM}\rightarrow W W}}=\left[\frac{v_1}{v_w} c_i+2 \frac{v_2}{v_w} d_i\right]^2\equiv \hat\Gamma^W_i\,,\label{GW}\\
\frac{\sigma_{q\bar{q}\rightarrow Z h_i}}{\sigma_{q\bar{q}\rightarrow Z h_{SM}}}&=&\frac{\sigma_{q q\rightarrow q q h_i}}{\sigma_{q q\rightarrow q q h_{SM}}}=\frac{\Gamma_{h_i\rightarrow Z Z}}{\Gamma_{h_{SM}\rightarrow Z Z}}=\left[\frac{v_1 c_i}{v_w+2 v_1/v_w} +\frac{4 v_2 d_i}{v_w+2 v_1/v_w}\right]^2 \equiv \hat\Gamma^Z_i\,,\label{GZ}\\
\frac{\sigma_{g g\rightarrow h_i}}{\sigma_{g g\rightarrow h_{SM}}}&=&\frac{\sigma_{g g\rightarrow q\bar{q} h_i}}{\sigma_{g g\rightarrow q\bar{q} h_{SM}}}=\frac{\Gamma_{h_i\rightarrow g g}}{\Gamma_{h_{SM}\rightarrow g g}}=\frac{\Gamma_{h_i\rightarrow f \bar{f}}}{\Gamma_{h_{SM}\rightarrow f \bar{f}}}=\frac{v^2_w}{v^2_1} c^2_i\equiv \hat\Gamma^f_i\,,\label{Gf}
\eea
where $q^\prime$ represents a quark with weak isospin different from $q$. 

The calculation of the Higgs decay rate to photons is more involved. By adapting to MSCT the formulas given in \cite{Gunion:1989we} we can write
\be
\Gamma_{h_i\rightarrow \gamma\gamma}=\sum_j \frac{\alpha_e^2 m_{h_i}^3}{256 \pi^3 v_w^2}\left| N_c e_j F_{i,j} \right|^2,
\ee
where the index $j$ is summed over the MSCT heavy charged particles (basically the top quark and the $W$ boson, the charginos $E$ and $N$, plus the charged composite scalars) and the factors $F_{i,j}$ are defined by
\bea
F_{i,W}&=&\left[2+3 \tau_{i,W}+3 \tau_{i,W}\left( 2-\tau_{i,W} \right) f(\tau_{i,W})\right] \sqrt{\hat\Gamma^W_i}\,,\nonumber\\
F_{i,f}&=&-2 \tau_{i,f}\left[1+\left( 1-\tau_{i,f} \right) f(\tau_{i,f})\right] \sqrt{\hat\Gamma^f_i}\,,\nonumber\\
F_{i,h}&=&\tau_{i,h}\left[ 1-\tau_{i,h}  f(\tau_{i,h})\right] \lambda_i ,\ \, \tau_{i,j}=\frac{4 m_j^2}{m_{h_i}^2}\,,
\eea
with
\bea
f(\tau_{i,j})=\left\{
\begin{array}{ll}  \displaystyle
\arcsin^2\sqrt{1/\tau_{i,j}} & \tau_{i,j}\geq 1 \\
\displaystyle -\frac{1}{4}\left[ \log\frac{1+\sqrt{1-\tau_{i,j}}}
{1-\sqrt{1-\tau_{i,j}}}-i\pi \right]^2 \hspace{0.5cm} & \tau_{i,j}<1
\end{array} \right. ,
\label{eq:ftau}
\eea
and $\lambda_i$ defined by Eq.\eqref{HlinL}.

\begin{figure}[h]
\begin{center}
\includegraphics[width=19pc]{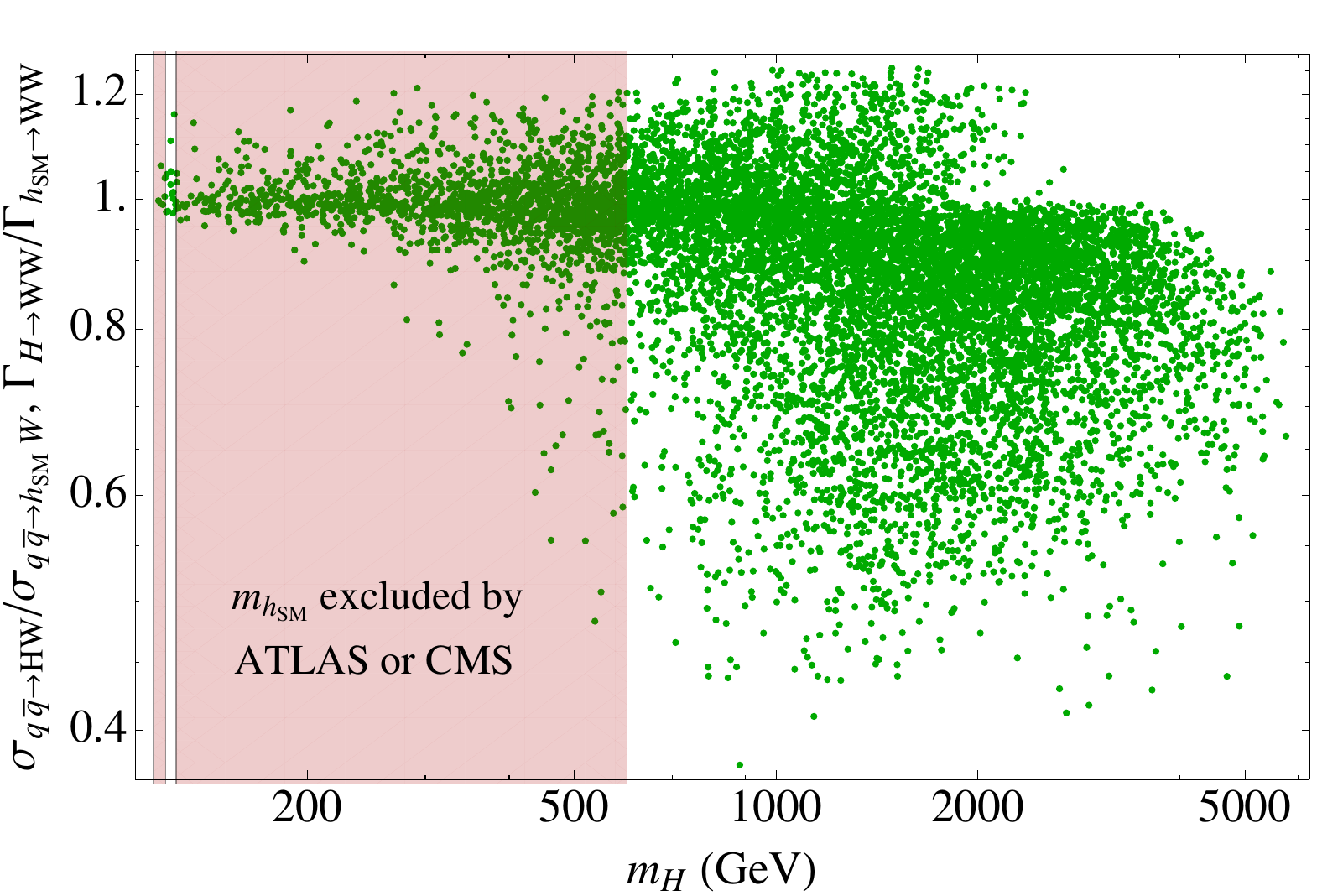}
\includegraphics[width=19.5pc]{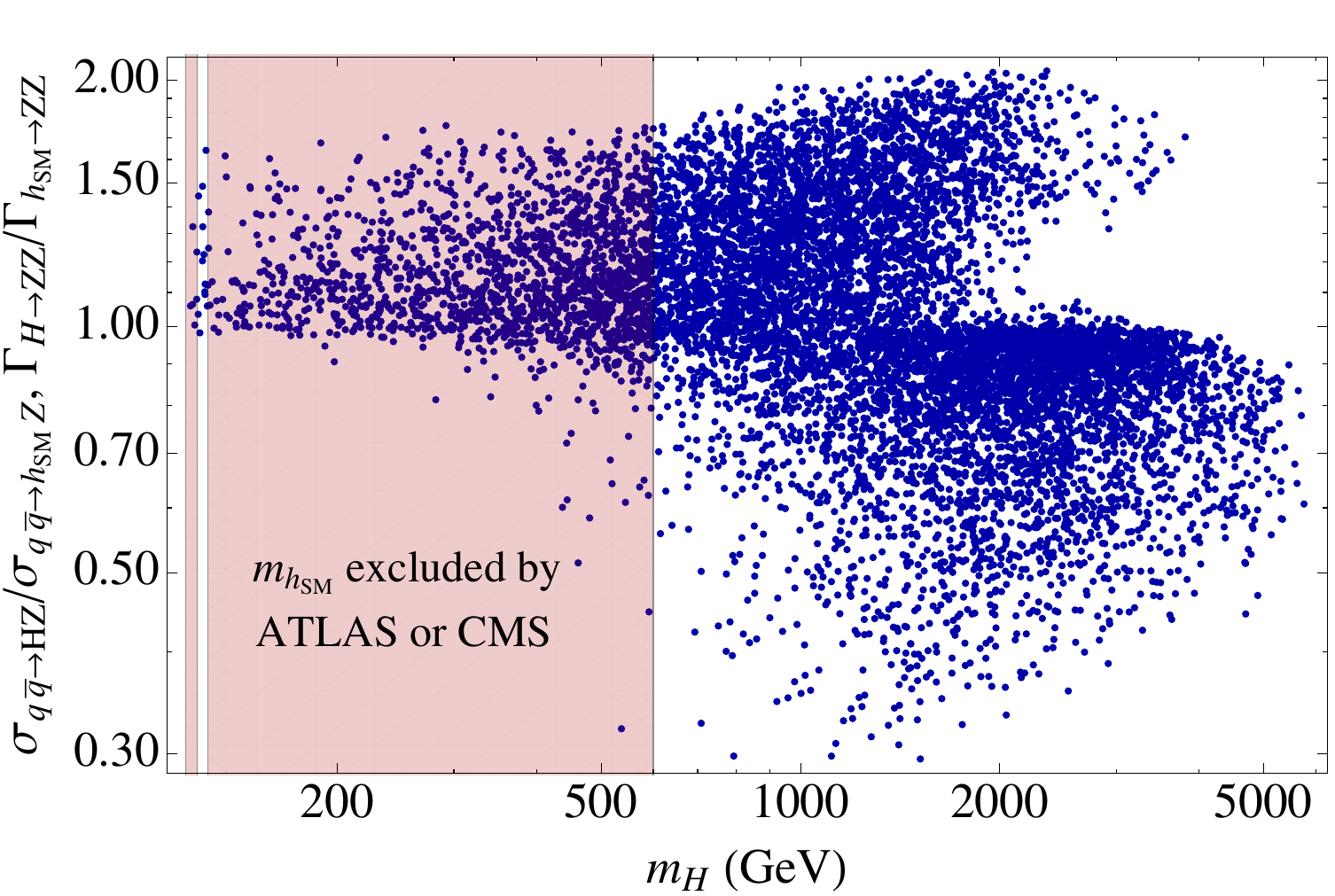}
\caption{\label{fig:HtoZZ} On the left (right) panel is shown the ratio between the decay rates to $W W$ ($Z Z$) of the MSCT neutral Higgs with the largest coupling to $W W$ and of the SM Higgs. The red bands show the range of $m_{h_{SM}}$ excluded either by ATLAS or CMS \cite{ATLAS-CONF-2012-019,CMS PAS HIG-12-008}.}
\vspace{1pc}
\end{center}
\end{figure}

\begin{figure}[h]
\begin{center}
\includegraphics[width=19pc]{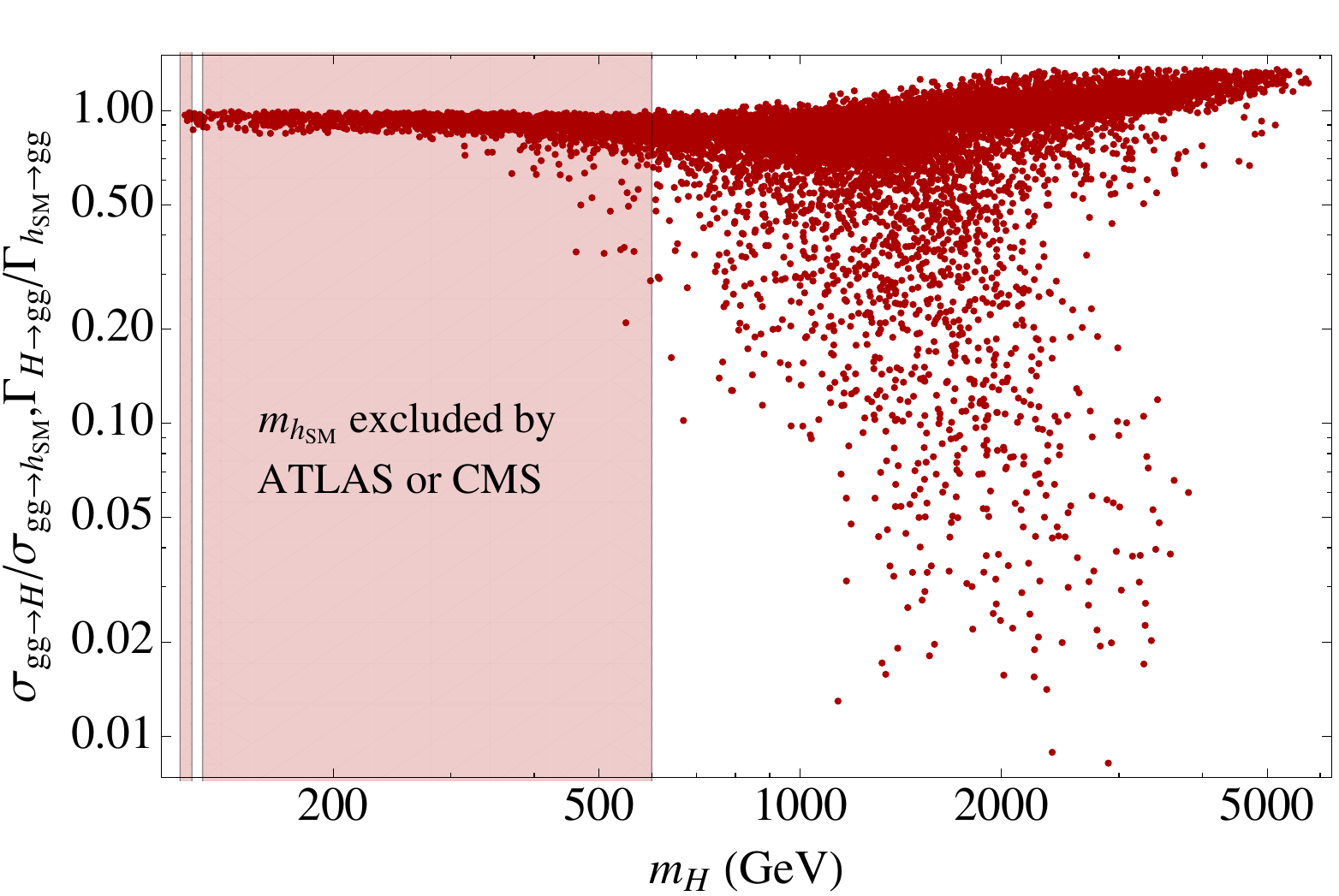}
\includegraphics[width=19pc]{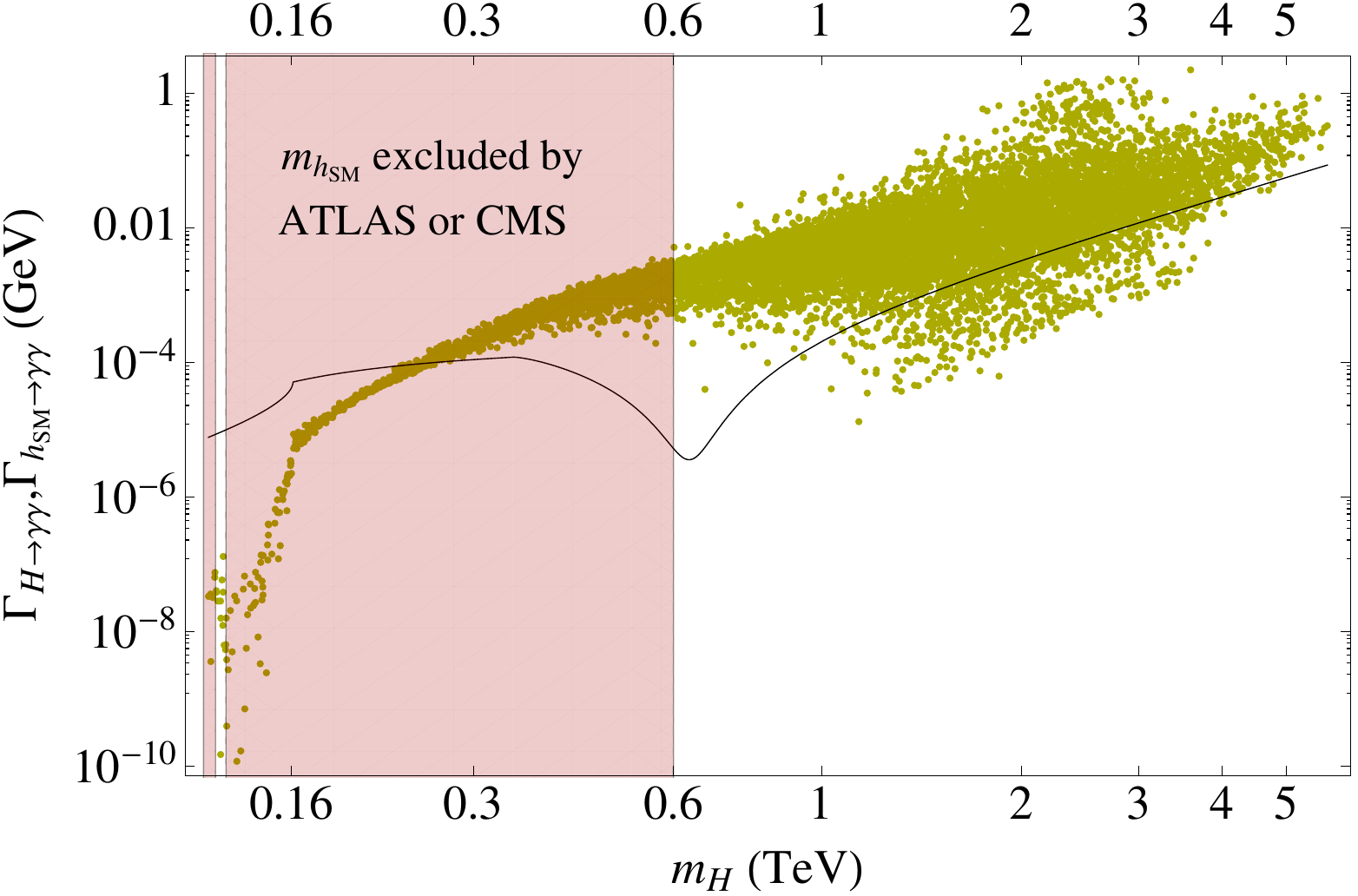}
\caption{\label{fig:ggtoH} Left panel: ratio between the production rate via gluon fusion of the MSCT neutral Higgs with the largest coupling to $W W$, and the corresponding rate for the SM Higgs. Right panel: decay rate to $\gamma\gamma$ of the MSCT neutral Higgs with the largest coupling to $W W$ (yellow dots) and SM Higgs (black line). The red bands show the range of $m_{h_{SM}}$ excluded either by ATLAS or CMS \cite{ATLAS-CONF-2012-019,CMS PAS HIG-12-008}.}
\vspace{1pc}
\end{center}
\end{figure}

Among the three neutral composite Higgs scalars available in MSCT we pick the one with the strongest coupling to $W$, which we label simply by $H$. The neutral scalar $H$ corresponds for most of the data points to either $h_1$ or $h_2$, and in the remaining cases to $h_3$. In Fig.\eqref{fig:HtoZZ} we show the numerical values of $\hat\Gamma^W_H$ (left panel) and $\hat\Gamma^Z_H$ (right panel), defined respectively in Eqs.(\ref{GW},\ref{GZ}) (where $i$ for the state $H$ is the one having the largest value of  $\hat\Gamma^W_i$), for each of the $10^4$ viable points defined in the previous section. For the same sample of points we show in Fig.\eqref{fig:ggtoH} the numerical values of $\hat\Gamma^f_H$ (left panel) and the $H$ decay rate to two photons (right panel) for both MSCT (yellow dots) and SM (black line). 

From Figs.(\ref{fig:HtoZZ},\ref{fig:ggtoH}) it is clear that the total decay rate $\Gamma_{tot}$ of $H$ for $m_H<1$ TeV is of the same order as that of the SM Higgs, which is equal roughly to 1 TeV for $m_{h_{SM}}=1$ TeV \cite{Djouadi:2005gi}. Since $\Gamma_{tot}$ is of $O(m_H)$ when $m_H>1$ TeV the narrow width approximation cannot be used to calculate the cross section of processes involving the production and decay of a Higgs from the Higgs production and decay rates, and a full calculation of the cross section would be necessary to study the Higgs phenomenology at LHC for $m_H>1$ TeV. For $m_H<1$ TeV we find that $H$ is mostly equal to the lightest neutral Higgs scalar, and therefore the $H$ decays to other scalars and pseudoscalars, mostly off shell, have a negligible branching ratio. The contribution to $\Gamma_{tot}$ of the charginos $N$ and $E$ is expected to be an order of magnitude smaller than $\Gamma_{H\rightarrow W W}$. From Figs.(\ref{fig:HtoZZ},\ref{fig:ggtoH} (left panel)) it is clear that for $m_H\lesssim 500$ GeV the $H$ production rate, accounted for mostly by the two processes in Fig.(\ref{fig:Hprod}) with a gluon-gluon initial state, is almost identical in value to that of the SM Higgs. Since for the same mass range the $H$ decay rate to $W W$ is roughly the same of that of the SM Higgs, while the $H$ decay rate to $ZZ$ is in general greater than the corresponding SM value, we expect most of the MSCT parameter space featuring $127.5< m_H/ {\rm GeV}< 500$ and $m_H/ {\rm GeV}< 122.5$ GeV to be ruled out by the 4.6 to 4.9 ${\rm fb}^{-1}$ ATLAS or CMS results, as it is the case for $h_{SM}$ \cite{ATLAS-CONF-2012-019,CMS PAS HIG-12-008}.

From Fig.(\ref{fig:ggtoH}) (right panel) it is clear that a light $H$ decay to two photons is disfavored compared to the SM case, while Fig.(\ref{fig:ggtoH}) (left panel) shows that the $H$ decay rate to $b\bar{b},\tau\bar{\tau}$, and $gg$, are almost identical in the low mass range to that of the SM Higgs. 
In this section we considered the case of relatively heavy composite Higgses since the points featuring $m_{h_1}<145$ GeV are about 0.5 \%. However, reducing the SUSY breaking scale to be of the same order of $\Lambda_{TC}$, while still breaking the EW dynamically,  light elementary scalar fields emerge more naturally. This is the scenario that we study in the following section.

\section{Keeping the Higgses Light}
\label{LSSs}
In this section we study a viable scenario in which
$m_{SUSY}=\Lambda_{TC}$ and we do not integrate out the elementary
Higgses coming from the MSSM sector.  We have taken positive mass squared for the elementary Higgses  so that EW breaking is triggered by the dynamical sector. In fact, the Yukawa term in Eq.\eqref{spmwt} coupling $\tilde{H}_u$ to the technifermions, once these condense in the infrared, generates a linear term in $\tilde{H}_u$ that induces a nonzero scalar VEV. The fact that this happens for generic values of $\mu,m_u,m_d,$ and $b$ is clearly a major difference with MSSM, for which in particular $\mu$ is constrained to be of $O(v_w)$. 

Another important consequence of the lack of EW symmetry breaking constraints on $\mu,m_u,m_d,$ and $b$, is that the tree-level Higgs mass is not bounded from above by $m_Z$ and can alone satisfy the experimental constraints. Consequentially the amount of fine tuning in MSCT with light elementary Higgs scalars is  reduced compared to MSSM, for which a large portion of the physical Higgs mass has to be generated through quantum corrections.
 
The low energy effective Lagrangian including light elementary Higgs scalars at the EW scale, below $\Lambda_{TC}$, was derived in \cite{Antola:2010nt}. The ETC-like induced sector is:
\be
{\cal L}_{ETC}^{\prime }=-c_1\Lambda^2_{TC} {\rm Tr}\left[ M \Delta \right]-c_5\Lambda^2_{TC} {\rm Tr}\left[ M J \right] + h. c.,
\label{eLETC}
\ee
where the matrix of the elementary up-type Higgs field $J$ is
\begin{eqnarray}
J & = &\frac{y_U}{\sqrt{2}}\left(\begin{array}{ccccc}
0 & 0 & \tilde{H}^0_{u} & 0\\
0 & 0 & -\tilde{H}^+_u & 0\\
\tilde{H}^0_u & -\tilde{H}^+_u & 0 & 0\\
0 & 0 & 0 & 0\\
\end{array}\right).
\end{eqnarray}
The Yukawa couplings of the elementary Higgs scalars to leptons and quarks remain unchanged with respect to the microscopic case given in Eq~\eqref{Lmic}:
\bea
-\mathcal{L}^\prime_{\text{Y,MSCT}} & = & \tilde{H}_u \cdot F^\prime_u + \tilde{H}_d \cdot  F_d+\textrm{h.c.} \\
F^\prime_u & = & q_{Lu}^i Y_u ^{i}\bar{u}_R^i+y_N L_L \bar{N}_R\, ,
\eea
with $F_d$ defined in  Eq~\eqref{Lmic}.
The complete Lagrangian is defined by
\be
{\cal L}_{MSCT}^{\prime \prime}={\cal L}_{SM kin}+{\cal L}_{MWT} + {\cal L}_{ETC}^{\prime}\label{HfulL}+\mathcal{L}^\prime_{\text{Y,MSCT}}\  .
\ee
The VEV of $M$ is still given by Eq.(\ref{Mvev}), while
\be
\langle \tilde{H}_u^0 \rangle=\frac{v_H s_\beta}{\sqrt{2}} \  ,\qquad \qquad  \langle \tilde{H}_d^0 \rangle=\frac{v_H c_\beta}{\sqrt{2}} \ .
\label{hvev}
\ee
The VEVs above are triggered by the TC condensate via the mixing in Eq.~\eqref{eLETC}.  In the previous section we have shown that $v_2$ is small because of its contribution to $T$, and therefore to simplify the present analysis we take $v_2=0$. 

The equations defining the parameters values that minimize the scalar potential are:
\bea
m_u^2&=& -\mu ^2+b~ \text{t}^{-1}_{\beta }+\frac{2 \text{cy}\, \text{s}^{-1}_{\beta } v_1}{v_H}+\frac{1}{8} c_{2 \beta } \left(g_L^2+g_Y^2\right) v_H^2,\  m_d^2= -\mu ^2+b~ t_{\beta   }-\frac{1}{8} c_{2 \beta } \left(g_L^2+g_Y^2\right) v_H^2,\ \nonumber \\
m_M^2&=& \frac{\lambda  v_1 \left(v_1^2-2 v_4^2\right) \left(v_1^2+v_4^2\right)-m_r^2 v_1^3+4 \text{cy}
   s_{\beta } v_4^2 v_H}{v_1^3-2 v_1 v_4^2},\ \lambda ''= 0,\ \lambda '=\frac{m_r^2 v_1-2 \text{cy} s_{\beta } v_H}{4 v_1 v_4^2-2 v_1^3},
\label{vevsc}
      \eea
with
\bea
\text{cy}=c_5\Lambda_{TC}\frac{y_U}{\sqrt{2}}.
\eea
For the VEV defined by Eqs.(\ref{Mvev},\ref{vevc},\ref{hvev}) the $W^\pm$ and $Z$ bosons squared masses are 
\be
m_Z^2=\frac{1}{4} \left(g_L^2+g_Y^2\right) \left(v_1^2+v_H^2\right),\ m_{W^\pm}^2=\frac{1}{4} g_L^2 \left(v_1^2+v_H^2\right).
\ee
The masses of the upper component $u$ and the lower one $d$ of a generic EW fermion doublet are given by
\be
m_u= y_u \frac{v_H}{\sqrt{2}} ,\ m_d=\frac{y_d}{y_u}t^{-1}_{\beta} m_u\ .
\ee
The derivation of the scalar squared mass matrices is straightforward, though the results are rather lengthy, and for this reason we do not list them here.

We repeat the scan of the parameter space but with $m_{SUSY} = \Lambda_{TC}$ to study the features of the mass spectrum in the phenomenological viable region. We use the same range for the independent parameters given in Eqs.(\ref{scanR}), except for $v_1$ that we allow to be as small as 100 GeV. We do not allow smaller values of $v_1$ so that $\Lambda_{TC}=g v_1$ remains close to the TeV scale. We also take $c_5=1/x$ and $x=4\pi$.  This choice of parameters means that the TC as well as the SUSY scales can be as low as $4\pi 100$ GeV$\simeq 1.2$ TeV. To simplify the particle spectrum we take $\mu$, which gives the mass of the Higgsinos, equal to $\Lambda_{TC}$, so that we are allowed to neglect those fields while having in principle still light elementary Higgs mass parameters. We furthermore require a zero VEV to be a stable minimum of the fundamental Higgs potential, so that a nonzero $v_H$ can be generated only by the techniquark condensate. We therefore impose on the soft parameters of the fundamental Higgs sector the constraints
\be
\left(m_u^2+\mu ^2\right) \left(m_d^2+\mu ^2\right)>b^2,\ m_u^2>0,\ m_d^2>0\ .
\ee

\begin{figure}[h]
\begin{center}
\includegraphics[width=20pc]{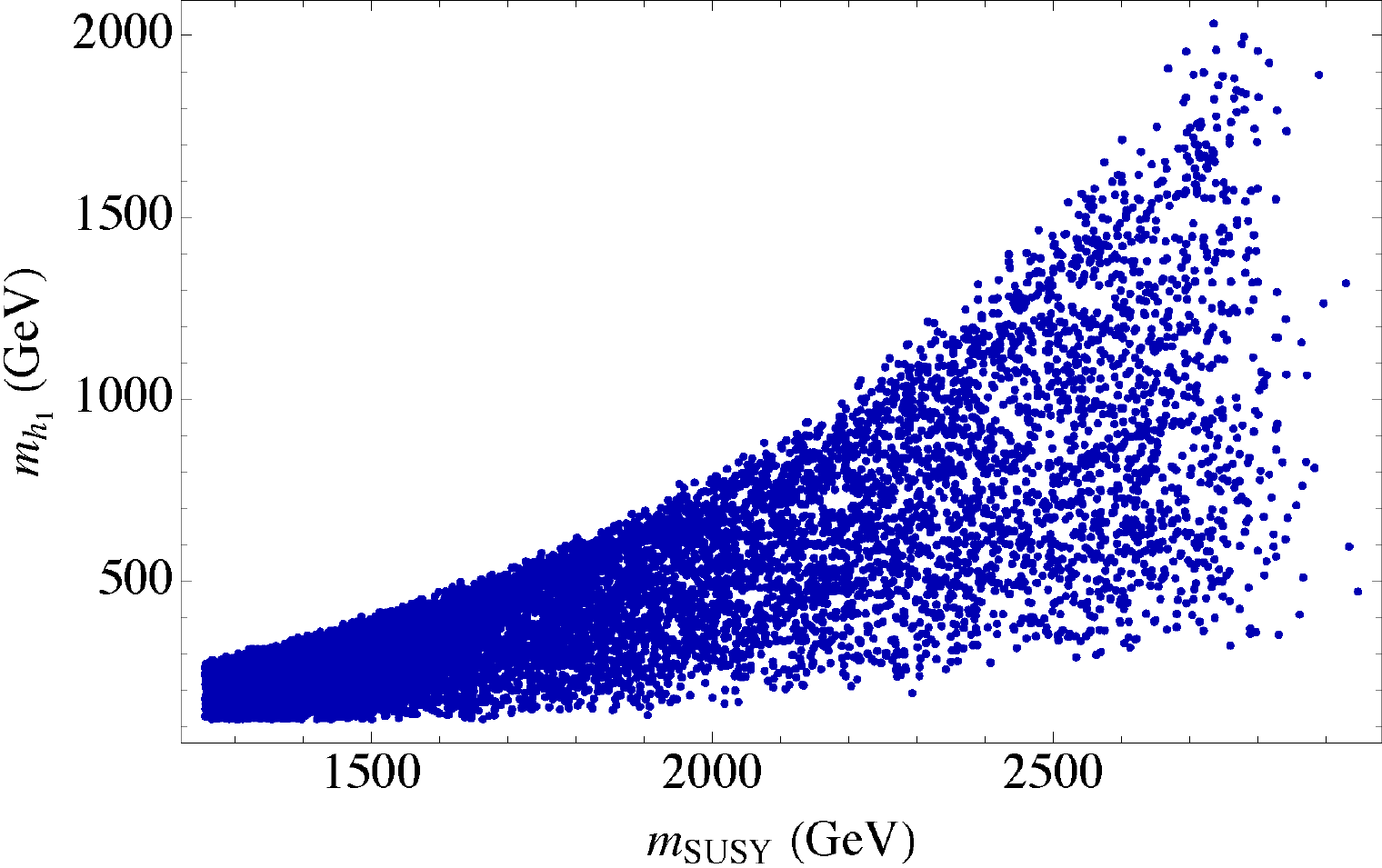}
\caption{\label{mSmH} Dependency of the mass of the lightest neutral scalar $h$, which is now a mixture of composite and elementary fields, on the SUSY breaking scale.}
\vspace{1pc}
\end{center}
\end{figure}

We stopped the scan of the parameter space once we have collected 10000 viable points.  In Fig.(\ref{mSmH}) we show the dependency of the tree-level mass of the lightest neutral scalar $h_1$, which is now a mixture of composite and elementary fields, on the SUSY breaking scale. For $m_\textrm{SUSY}$ around 1.2 TeV $h_1$ is shown to be light compared to the EW scale. A light neutral Higgs scalar appears  because of the mixing with a fundamental scalar field 
\cite{Antola:2009wq}. The fact that MSCT is able to achieve a tree-level Higgs mass large enough to satisfy the experimental constraints shows clearly that the fine tuning of the model is greatly reduced with respect to that of the MSSM with a comparable SUSY breaking scale.

The phenomenology of MSCT in the regime with light fundamental scalars is expected to be very rich, with signals associated to MSSM superpartners with masses of $O(\Lambda_{TC})$ as well as other massive composite states, such as technivectors. We plan, in the future, to add the spin one states due to the TC dynamics following Ref.~\cite{Foadi:2007ue}.

\section{Conclusions}
\label{conclusione}
We provided a complete extension of MWT able to account for the SM fermion masses. The extension becomes supersymmetric at energies greater or equal to the TC compositeness scale and we dubbed it Minimal Super Conformal Technicolor (MSCT). We investigated this model in two main distinct  regimes: one in which all the elementary scalars are heavy compared to the TC compositeness scale and another in which the elementary Higgs scalars are kept light with respect to the same scale. In the latter case we argued that the fine tuning of the model is dramatically reduced compared to that of MSSM, since the tree-level Higgs mass in MSCT is able alone to accommodate the experimental bounds. In both cases we showed, by performing a large parameter scan, that the model is phenomenologically viable. We also determined that MSCT does not require, contrarily to MSSM, the $\mu$ parameter to be of $O(v_w)$. Furthermore we studied the composite Higgs phenomenology at the LHC for the case in which the elementary Higgses are heavy. We have then compared the results to the SM Higgs phenomenology and shown that it is possible, depending on the parameter space, to disentangle the two models. LHC already constrains severely the low energy portion of the model's parameter space.   

Our model can be viewed as a working complete example of an extended TC theory able to give masses to the SM fermions. One of the main results of our work is that the extension back reacts heavily on the TC dynamics, vacuum and spectral properties. This feature is expected to hold for large classes of models of dynamical electroweak symmetry breaking.
 
 \acknowledgments
 We thank R. S. Chivukula and E. H. Simmons for useful discussions and M. Nardecchia for helpful comments. 

\newpage
\appendix

\section{$\mathcal{N}=4$ Lagrangian}
\label{appn4}

We first define the fields $\varphi_{ij}^{a}$ and $\eta_{i}^{a}$. Here $a$ is the SU$(2)$ adjoint index which is omitted for clarity in the following definitions:

\begin{eqnarray}
\phi_{i} & = & (\tilde{U}_{L},\tilde{D}_{L},\bar{\tilde{U}}_{R})_{i}\;,\;\varphi_{i4}=\frac{1}{2}\phi_{i}\;,
\;\varphi_{ij}=\frac{1}{2}\epsilon_{ijk}\bar{\phi}_{k}\;,\;\varphi^{T}=-\varphi\end{eqnarray}
\be
\eta_{i}=(U_{L},D_{L},\bar{U}_{R},\bar{D}_{R})_{i} \;.
\ee
The fields transform under $g\in$SU$(4)_{R}$ as $\varphi\rightarrow 
g\varphi g^{T}$ and $\eta\rightarrow g\eta$. The Lagrangian of the $\mathcal{N}=4$ sector is then:

\begin{eqnarray}
\mathcal{L} & = & -\frac{1}{4}G^{\mu\nu a}G_{\mu\nu}^{a}-\text{Tr}(D^{\mu}\bar{\varphi})^{a}(D_{\mu}\varphi)^{a}-i\bar{\eta}^{a}\bar{\sigma}^{\mu}(D_{\mu}\eta)^{a}\\
 &  & +\sqrt{2}gf^{abc}\left(\eta^{Ta}\bar{\varphi}^{b}\eta^{c}+\text{c.c.}\right)\label{n4yukawa}\\
 &  & -\frac{1}{2}g^{2}\left(f^{abd}f^{ace}+f^{abe}f^{acd}\right)\text{Tr}\left(\bar{\varphi}^{b}\varphi^{c}\right)\text{Tr}\left(\bar{\varphi}^{d}\varphi^{e}\right)
\end{eqnarray}

\section{Scalar Mass Matrices}
\label{AppA}

The components of the matrix $M\sim \eta^T \eta$ can be described either in terms of wave functions of the underlying techniquarks or the transformation properties of the composites under $\text{SU}(2) \times \text{U}(1)$. In terms of the latter description, we have the states listed in Table (\ref{Mfields}).

\begin{table}[h]
\centering
\begin{tabular}{c|c|c}
\noalign{\vskip\doublerulesep}
Field & SU$(2)_{\text{L}}$ & U$(1)_{\text{Y}}$ \tabularnewline[\doublerulesep] \hline \noalign{\vskip\doublerulesep}
$\Delta\sim Q_L Q_L$ & $\square \square$ & 1 
\tabularnewline[\doublerulesep] \noalign{\vskip\doublerulesep}
$\sigma_U \sim Q_L \bar{U}_R$ & $\square$ & $-\frac{1}{2}$
\tabularnewline[\doublerulesep] \noalign{\vskip\doublerulesep}
$\sigma_D \sim Q_L \bar{D}_R$ & $\square$ & $\frac{1}{2}$
\tabularnewline[\doublerulesep] \noalign{\vskip\doublerulesep}
$\delta^{++} \sim \bar{U}_R \bar{U}_R$ & 1 & 2
\tabularnewline[\doublerulesep] \noalign{\vskip\doublerulesep}
$\delta^+ \sim  \bar{U}_R  \bar{D}_R$ & 1 & 1
\tabularnewline[\doublerulesep] \noalign{\vskip\doublerulesep}
$\delta^0 \sim  \bar{D}_R  \bar{D}_R$ & 1 & 0
\tabularnewline[\doublerulesep] \end{tabular}\caption{Component fields of the matrix $M$ in terms of $\text{SU}(2) \times \text{U}(1)$ transformation properties.}
\label{Mfields}
\end{table}

In this notation, we have
\be
 M=\left(\begin{array}{cccc}
\sqrt{2}\Delta^{++}
& \Delta^+
& \sigma_U^0
& \sigma_D^+
\\
\Delta^+
& \sqrt{2}\Delta^0
& \sigma_U^{+*}
& \sigma_D^0
\\
\sigma_U^0
& \sigma_U^{+*}
& \sqrt{2}\delta^{++}
& \delta^+
\\
\sigma_D^+
& \sigma_D^0
& \delta^+
& \sqrt{2}\delta^0
\end{array}\right)\label{Matrix}
\ee
where all fields are complex valued. When the electroweak symmetry is spontaneously broken, these states mix, and we now define the language with which these states are referred to.

We define the following mixing matrices so that for each type of scalar the lower index is ordered by increasing mass. The neutral scalars and pseudoscalars mixing matrices are defined as follows:
\be
\left(\begin{array}{c}
 h_1 \\
 h_2 \\
 h_3 \\
\end{array}\right)
=
U_s\frac{\Re}{\sqrt{2}} 
\left(\begin{array}{c}
    \sigma_U^0\\
 \Delta^0 \\
  \delta^0 \\
\end{array}\right)
\quad,\quad 
\left(\begin{array}{c}
  G^0 \\
 A_1 \\
 A_2 \\
\end{array}\right)
=
U_p\frac{\Im}{\sqrt{2}} 
\left(\begin{array}{c}
   \sigma_U^0\\
  \Delta^0 \\
 \delta^0 \\
\end{array}\right)
\label{mixM}
\ee
where $G^0$ is now the would-be Goldstone. The states $\phi=\Re \sigma_D^0/\sqrt{2}$ and $A_\phi=\Im \sigma_D^0/\sqrt{2}$ do not mix. Correspondingly, for the charged states we write
\be
\left(\begin{array}{c}
 G^+ \\
 h^+_1 \\
 h^+_2 \\
h^+_3 \\
\end{array}\right)
= 
U_+
\left(\begin{array}{c}
 \sigma_U^+ \\
 \sigma_D^+ \\
 \Delta^+ \\
 \delta^+ \\
\end{array}\right) 
\quad,\quad
\left(\begin{array}{c}
 h^{++}_1 \\
 h^{++}_2 \\
\end{array}\right)
= 
U_{++}
\left(\begin{array}{c}
 \Delta^{++} \\
 \delta^{++} \\
\end{array}\right) 
\ee
where $G^+$ is the would-be Goldstone.

The squared mass matrix for the neutral scalars above, prior to diagonalization, is
\be
{\cal M}^2_{h}= 2 \left(
\begin{array}{ccc}
 v_1^2 \left(\lambda +2 \lambda '\right) & v_1 \left(\lambda  v_2+2 v_4 \lambda ''\right) & v_1 \left(\lambda  v_4+2 v_2 \lambda ''\right) \\
 v_1 \left(\lambda  v_2+2 v_4 \lambda ''\right) & v_2^2 \left(\lambda +4 \lambda '\right)-\frac{v_1^2 v_4 \lambda ''}{v_2} & \lambda  v_2 v_4+v_1^2 \lambda '' \\
 v_1 \left(\lambda  v_4+2 v_2 \lambda ''\right) & \lambda  v_2 v_4+v_1^2 \lambda '' & v_4^2 \left(\lambda +6 \lambda '\right)-2 v_2^2 \lambda '-\frac{v_1^2 v_4 \lambda ''}{v_2} 
\end{array}
\right).
\ee
The squared mass matrix for the neutral pseudoscalars is:
\be
{\cal M}^2_A= -2 \left(
\begin{array}{ccc}
 4 v_2 v_4 \lambda '' & 2 v_1 v_4 \lambda '' & 2 v_1 v_2 \lambda '' \\
 2 v_1 v_4 \lambda '' & \frac{v_1^2 v_4 \lambda ''}{v_2} & v_1^2 \lambda '' \\
 2 v_1 v_2 \lambda '' & v_1^2 \lambda '' & 2 \left(v_2^2-v_4^2\right) \lambda '+\frac{v_1^2 v_4 \lambda ''}{v_2} 
\end{array}
\right).
\ee
The squared mass matrix for the charged scalars is:
\bea
&&{\cal M}^2_{h^\pm}= 2\\
&&\left(
\begin{array}{cccc}
 v_1^2 \left(\lambda '-\frac{v_4 \lambda ''}{v_2}\right) & \sqrt{2} v_1 \left(v_2 \lambda '-v_4 \lambda ''\right) & 0 & 0 \\
 \sqrt{2} v_1 \left(v_2 \lambda '-v_4 \lambda ''\right) & 2 v_2 \left(v_2 \lambda '-v_4 \lambda ''\right) & 0 & 0 \\
 0 & 0 & \left(v_1^2-2 v_2^2+2 v_4^2\right) \lambda '-\frac{v_1^2 v_4 \lambda ''}{v_2} & \sqrt{2} v_1 \left(v_4 \lambda '-v_2 \lambda ''\right) \\
 0 & 0 & \sqrt{2} v_1 \left(v_4 \lambda '-v_2 \lambda ''\right) & \left(v_1^2-2 v_2^2+2 v_4^2\right) \lambda '-\frac{v_1^2 v_4 \lambda ''}{v_2}
\end{array}
\right). \nonumber
\eea
The squared mass matrix for the doubly charged scalars is:
\be
{\cal M}^2_{h^{\pm\pm}}=2 \left(
\begin{array}{cc}
 -2 v_2^2 \lambda '+v_1^2 \left(2 \lambda '-\frac{v_4 \lambda ''}{v_2}\right) & v_1^2 \lambda '-2 v_2 v_4 \lambda '' \\
 v_1^2 \lambda '-2 v_2 v_4 \lambda '' & -2 v_2^2 \lambda '+v_1^2 \left(2 \lambda '-\frac{v_4 \lambda ''}{v_2}\right)
\end{array}
\right).
\ee
 The neutral composite scalar $\phi$ and pseudoscalar $A_\phi$ respectively having the quantum numbers of the real and imaginary part of the bound state $ \sigma^0_D$  have squared masses
\be
m^2_{\phi}=2 \left(v_4 + v_2\right) \left(2 v_4 \lambda '-\frac{v_1^2 \lambda ''}{v_2}\right)\,,\quad m^2_{A_\phi}= 2 \left(v_4-v_2\right) \left(2 v_4 \lambda '-\frac{v_1^2 \lambda ''}{v_2}\right) \ . 
\label{mh4}
\ee
Besides the massless Goldstone boson eaten by the $Z$ boson, the physical particle spectrum features two more neutral pseudoscalars with squared masses
\bea
m^2_{A_{1,2}}&=&\frac{1}{2} \left(\left(v_4^2-v_2^2\right) \lambda '-\frac{v_1^2 v_4 \lambda ''}{v_2} -2 v_2 v_4 \lambda '' \right.\nonumber \\ 
& \mp & \left. \sqrt{\left(v_2^2-v_4^2\right)^2 \lambda '^2+4 v_2 v_4 \left(v_4^2-v_2^2\right) \lambda ' \lambda ''+\left(v_1^4+4 v_2^2 \left(v_1^2+v_4^2\right)\right)
   \lambda ''^2}\right).
\label{mpi}
\eea

The squared masses of the charged physical scalars are:
\be
m^2_{h^\pm_1}=2 \left(v_1^2+2 v_2^2\right) \left(\lambda '-\frac{v_4 \lambda ''}{v_2}\right),\ m^2_{h^\pm_{2,3}}=4 \left(v_4^2-v_2^2\right) \lambda '+2 v_1^2 \left(\lambda '-\frac{v_4 \lambda
   ''}{v_2}\right)\mp 2 \sqrt{2} \left|v_4 \lambda '-v_2 \lambda ''\right| v_1.
\label{mhp}
\ee
The squared masses of the doubly charged physical scalars are:
\be
m^2_{h^{\pm\pm}_1}=2 \left(v_1^2-2 v_2^2\right) \left(\lambda '-\frac{v_4 \lambda ''}{v_2}\right),\ m^2_{h^{\pm\pm}_2}=2 v_1^2 \left(3 \lambda '-\frac{v_4 \lambda ''}{v_2}\right)-4 v_2 \left(v_2 \lambda '+v_4 \lambda
   ''\right)
\label{mhpp}
\ee

\end{document}